\documentclass[aps,prd,eqsecnum,showpacs,superscriptaddress,
twocolumn,nofootinbib,floatfix,preprintnumbers,altaffilletter]{revtex4}
\usepackage{graphicx}
\usepackage{hyperref}

\usepackage{amssymb}
\usepackage{amsmath}
\usepackage{longtable}
\usepackage{rotating}
\usepackage{color}
\usepackage{fancyhdr}

\def\lesssim{\mathrel{\hbox{\rlap{\hbox{\lower4pt\hbox{$\sim$}}}\hbox{$<$}}}}
\def\gtrsim{\mathrel{\hbox{\rlap{\hbox{\lower4pt\hbox{$\sim$}}}\hbox{$>$}}}}

\def\aap{\ref@jnl{A\&A}}                % Astronomy and Astrophysics
\def\CQG{\ref@jnl{Class. Quantum Grav.}}

\def\mnras{MNRAS}

\newcommand{\red}[1]{}

\newcommand{\be}{\begin{equation}}
\newcommand{\ee}{\end{equation}}
\newcommand{\bea}{\begin{eqnarray}}
\newcommand{\eea}{\end{eqnarray}}
\newcommand{\bdm}{\begin{displaymath}}
\newcommand{\edm}{\end{displaymath}}

\newcommand{\etal}{et al.}

\begin{document}
\title{The Gravitational-Wave Discovery Space of Pulsar Timing Arrays}
%DRAFT: NOT  FOR DISTRIBUTION}
\author{Curt Cutler}
\affiliation{Jet Propulsion Laboratory, California Institute of Technology, 4800 Oak Grove Dr., Pasadena, CA 91109}
\affiliation{Theoretical Astrophysics, California Institute of
	Technology, Pasadena, California 91125}
\author{Sarah Burke-Spolaor}
\affiliation{Jet Propulsion Laboratory, California Institute of Technology, 4800 Oak Grove Dr., Pasadena, CA 91109}
\author{Michele Vallisneri}
\affiliation{Jet Propulsion Laboratory, California Institute of Technology, 4800 Oak Grove Dr., Pasadena, CA 91109}
\affiliation{Theoretical Astrophysics, California Institute of
	Technology, Pasadena, California 91125}
\author{Joseph Lazio}
\affiliation{Jet Propulsion Laboratory, California Institute of Technology, 4800 Oak Grove Dr., Pasadena, CA 91109}
\author{Walid Majid}
\affiliation{Jet Propulsion Laboratory, California Institute of Technology, 4800 Oak Grove Dr., Pasadena, CA 91109}

%% for REVTeX4, each author name can be set in a separate \author{} field

%\draft

\date{\today}

\begin{abstract}
Recent years have seen a burgeoning interest in using pulsar timing arrays (PTAs) as gravitational-wave (GW) detectors. To date, that interest has focused mainly on three particularly promising source types: supermassive--black-hole binaries, cosmic strings, and the stochastic background from early-Universe phase transitions.
In this paper, by contrast, our aim is to investigate the PTA potential for discovering unanticipated sources.  We derive significant constraints on the available discovery space based solely on energetic and statistical considerations: we show that a PTA detection of GWs at frequencies above $\sim 3 \times 10^{-5}$ Hz would either be an extraordinary coincidence or violate ``cherished beliefs;''
we show that for PTAs GW memory can be more detectable than direct GWs, and that, as we consider events at ever higher redshift, the memory effect increasingly dominates an event's total signal-to-noise ratio.
The paper includes also a simple analysis of the effects of pulsar red noise in PTA searches, and a demonstration that the effects of periodic GWs in the $\sim 10^{-7.95}\mbox{--}10^{-4.5}\,$ Hz band would \emph{not} be degenerate with small errors in standard pulsar parameters (except in a few narrow bands).
\end{abstract}
\pacs{04.80.Nn, 04.25.dg, 04.30.-w, 95.85.Sz, 97.80.-d, 97.60.Gb} 
% Curt grabbed these pacs from related papers on 8/13/13
\maketitle %% NULL FUNCTION WITH LATEX 2e

\section{Introduction}
\label{Sec:Int}
The idea of detecting gravitational waves (GWs) by monitoring the arrival times of radio pulses from neutron stars (i.e., by \emph{pulsar timing}) was first proposed by Sazhin \cite{1978SvA....22...36S} and Detweiler \cite{1979ApJ...234.1100D}; its modern formulation by Hellings and Downs~\cite{1983ApJ...265L..39H} emphasizes the importance of searching for \emph{correlations} in the pulse-timing time deviations among an \emph{array} of intrinsically stable millisecond pulsars.
The last few years have seen a strong renewed interest in these
searches, with the formation of three major pulsar timing programs: the European Pulsar Timing Array (EPTA, \cite{2011MNRAS.414.3117V}),
% add \cite{Kramer2013} once it's out
the North American Nanohertz Observatory for Gravitational Waves (NANOGrav, \cite{2013ApJ...762...94D}),
% add \cite{McLaughlin2013} once it's out
and the Australian Parkes Pulsar Timing Array (\hbox{PPTA}, \cite{2013PASA...30...17M}),
% add \cite{Manchester2013b} once it's out
which have now joined into a global collaboration, the International Pulsar Timing Array (IPTA, \cite{2010CQGra..27h4013H}).
% \red{SBS: all CQG references need to be filled in once relevant papers are published or put on arXiv!}

The most promising \emph{known} sources of GWs for PTAs are in-spiraling supermassive black hole binaries (SMBHBs).  Some estimates suggest that these will be detected by PTAs as soon as $\sim$2016--2020~\cite{2013MNRAS.433L...1S}. The first detection could plausibly identify the inspiral waves from an orbiting SMBHB (see, e.,g.,\cite{2009MNRAS.394.2255S,2010MNRAS.407..669Y,2013arXiv1307.4086S}), the burst waves that follow its coalescence \cite{2010MNRAS.401.2372V,2012ApJ...752...54C} or a stochastic background from many SMBHBs (see, e.,g., \cite{2003ApJ...583..616J,2013MNRAS.433L...1S}). Pulsar timing already provides the most stringent upper limit on $\Omega_\mathrm{GW} \equiv \rho_\mathrm{GW}/\rho_0$ (the ratio of the energy density in GWs to the closure density of the Universe), and is beginning to impact standard theories of hierarchical structure formation via constraints on the SMBH merger rate.
% \red{add latest constraint and ref when available ~\citep{shannon-submitted} SBS: I hope that the shannon et al. limit will be released at our referee phase; we will quote that limit here then: an XX\% confidence limit of $\Omega_\mathrm{GW} < YYY$ in the band ZZZ.} 

In this article we explore the \emph{discovery} potential of PTAs.  Our main motivation is to minimize the risk that current observing strategies and planned data-analysis pipelines 
artificially preclude the discovery of various types of sources.
For instance, most pulsars in PTAs are currently observed with irregular cadences of $\sim 2\mbox{--}4$ weeks. The observational strategies for most pulsar timing arrays are currently optimized to be sensitive to the gravitational wave background (based on strategies as determined by \cite{2005ApJ...625L.123J}).
%\red{Curt doesn't understand the meaning of following phrase: (largely motivated by \cite{2005ApJ...625L.123J})} 
% SBS: I meant that the design of current PTAs are mostly based on that jenet paper, which determined how many pulsars/observations/RMS residuals were required to detect a GWB. I have tried to clarify this.
This is appropriate for GWs at the lowest observable frequencies  (of order the inverse of the total observation time, $\sim 10^{-8}$ Hz), where PTAs are particularly sensitive. However a search for GW bursts lasting (say) $10^5$\,s would clearly
benefit greatly from coordinated timing observations (using a few radio telescopes) that get repeated several times a day.
Thus we address the following questions:
\begin{itemize}
\item Is there a strong motivation for increasing the observing cadence to improve our sensitivity to GWs with frequencies $\sim 10^{-6}\mbox{--}10^{-5}$ Hz?   
% Is their detection hindered by errors in the pulsar timing-model parameters (such as sky position and spindown rate)? - MV: we repeat this question later
%
\item What constraints can we impose on the PTA discovery space 
based on simple energetic, statistical, and causality arguments? 
\end{itemize}

In addressing the first question, an important issue that arises is whether, even if strong sources exist in this band,
our sensitivity might be degraded by degeneracies between GW effects and small errors in the timing-model parameters of the monitored pulsars.
In addressing the second question, we are necessarily retracing some of the
trails blazed by Zimmermann and Thorne~\cite{Zimmermann:1982wi} (hereinafter ZT82) in their classic paper, ``The gravitational waves that bathe the Earth: upper limits based on theorists' cherished beliefs.''  However there are important differences between our paper and theirs:  
\begin{itemize}
\item ZT82 restricted attention to sources  at $z \alt 3$, while we consider the case of very high-$z$ sources as well.
\item Unlike ZT82, we include the ``memory effect'' among potential observables;  its detection turns out to be especially promising in the high-$z$ case.
\item ZT82 restricted attention to GWs in the frequency range $10^{-4} < f < 10^{4}\,$Hz (the band of interest for ground-based and space-based interferometers), while we focus on GWs with $f \alt 10^{-5} \,$Hz.
(However, there are several instances for which the ZT82 estimates extend trivially to lower frequency;  we will note these instances in our paper as they arise.)
\end{itemize}
%
% MV: what stochastic-background papers? GW-general, or just PTA, e.g.
% amplitude estimates: Phinney 2001astro.ph..8028P
%                      Jaffe 2003ApJ...583..616J
%                      Wyithe 2003ApJ...590..691W
%                      Sesana 2004ApJ...611..623S
%                      Sesana 2008MNRAS.390..192S
%                      Sesana 2013MNRAS.433L...1S
% limits:              Jenet 2006ApJ...653.1571J (Parkes)
%                      van Haasteren 2011MNRAS.414.3117V (EPTA)
%                      Demorest 2013ApJ...762...94D (NANOGrav)
% methods:             Jenet 2005ApJ...625L.123J
%                      van Haasteren 2009MNRAS.395.1005V
% detectability:       Yardley 2011MNRAS.414.1777Y
%                      Ravi 2012ApJ...761...84R
%

This paper is organized as follows. In Sec.\ \ref{sec:determ}
we describe a simple general framework for thinking about pulsar
timing observations, and we characterize how the detection
signal-to-noise ratio scales with quantities such as the number of
pulsars surveyed, the timing accuracy provided by each pulsar
observation, the observing cadence, and the total observation time.
We also briefly review pulsar timing noise, with some emphasis on its
red noise component.
In Sec.\ \ref{sec:source-review} we summarize salient results regarding PTA searches for SMBHBs and cosmic strings, largely to provide points of comparison with possible unknown GW sources.
In Sec.\ \ref{sec:degen} we demonstrate that the timing residual signatures of GWs in the $10^{-7.95}\mbox{--}10^{-4.5}\,$ Hz band are {\it not} degenerate with small errors in the pulsar parameters, except for very narrow frequency bands;
% at harmonics of $1/\mathrm{day}$ (
had this been otherwise, there would have been little point in considering more fundamental constraints on possible sources in this band.
In Secs.\ \ref{sec:z1} and \ref{sec:highredshift} we investigate what constraints on source strengths arise from fundamental considerations of energetics, statistics and causality.
In Sec.~\ref{sec:galactic} we discuss how are estimates get modified for highly beamed sources, and for sources in our Galaxy.
In Sec.\ \ref{sec:summ} we summarize our conclusions, listing some caveats.

Regarding notation, we adopt units in which $G=c=1$. Also, the signal frequency $f$, observation time $T_\mathrm{obs}$ and signal duration $T_\mathrm{sig}$ all refer to time as measured in the observer's frame, at the Solar system barycenter.

\section{The PTA signal-to-noise ratio for GW signals of known shape}
\label{sec:determ}

\subsection{Signal-to-noise ratio for white noise signals}

In the rest of this paper, we are going to assume an idealized,
general scaling law for the detection signal-to-noise ratio (SNR) of
an individual GW source, as observed by a pulsar timing array: to wit,
\newcommand{\snr}{\mathrm{SNR}}
\newcommand{\gw}{\mathrm{GW}}
\newcommand{\noise}{\mathrm{noise}}
\newcommand{\rms}{\mathrm{rms}}
\begin{equation}
\label{eq:snr}
%\snr^2 = \frac{\langle \delta t_\gw^2 \rangle M N}{\langle \delta t_\noise^2 \rangle},
\snr^2 =  M N \left\langle \frac{\delta t_\gw^2}{  \delta t_\noise^2} \right\rangle,
\end{equation}
where
\begin{itemize}
\item $\delta t_\gw$ is the \emph{timing residual} due to GWs;
\item $\delta t_\noise$ is the noise in the residuals,  which includes contributions from the observatory, from pulse propagation, and from intrinsic pulsar processes;
\item $\langle \cdots \rangle$ denotes the average over all pulsars in the PTA and over all observed pulses;
\item $M$ is the number of pulsars in the PTA; and
\item $N$ is the total number of observations for each pulsar.
\end{itemize}
In what follows, purely for simplicity we will assume that $\delta t_\rms$ is roughly the same across PTA pulsars and observations, so we define 
\be\label{replace}
\left\langle \frac{\delta t_\gw^2}{  \delta t_\rms^2} \right\rangle 
 = \frac{\langle \delta t_\gw^2 \rangle}{\delta t_\rms^2} \, .
\ee
with $\delta t_\rms$ a representative rms value for the noise.

The term ``timing residual'' requires definition: it is the difference
between the time of arrival (TOA) of a train of pulses \emph{observed}
at the radio telescope and the TOA \emph{predicted} by the best-fitting
\emph{timing model} for a pulsar. This deterministic model includes
parameters (such as the sky position of and distance to the pulsar) that
affect the propagation of signals to the observatory, as well as
parameters (such as the pulsar period and its derivatives and, if
needed, orbital elements for pulsars in binaries) that describe the
intrinsic time evolution of the pulsar's emission.

The pulses from millisecond pulsars are usually too weak to be
observed individually, so the TOAs refer to \emph{integrated} pulse
profiles obtained by ``folding'' the output of radiometers with the
putative pulsar period over observations with durations of tens of
minutes to an hour.  Typically, such pulsar timing observations are
repeated at intervals of two to four weeks, yielding sparse data sets;
however, the individual observations are often run
quasi-simultaneously at multiple receiving frequencies (typically one
hour to two days apart, since the feeds need to be switched), yielding a set of TOAs at the same \emph{epoch}. 
See \cite{2008LRR....11....8L,2003LRR.....6....5S} and references therein for more detail.

In analogy with other applications in GW data analysis \cite{maggiore2008}, our scaling for the SNR can be motivated by considering a ratio of \emph{likelihoods}: namely, the likelihood of the residual data $r_i$ (with $i$ indexing both epochs and pulsars) under the hypothesis that $r_i = g_i + n_i$, with $g_i$ describing a GW signal of known shape, and $n_i$ denoting noise; and the likelihood of the residuals under the noise-only hypothesis $r_i = n_i$. For Gaussian noise, when the GW signal is really present, the likelihood ratio is
\begin{equation}
\label{eq:likes}
\exp \, \{g_i (C^{-1})^{ij} g_j/2 + n_i (C^{-1})^{ij} g_j\}
\end{equation}
(summations implied), where $C_{ij} = \langle n_i n_j \rangle$ is the variance-covariance matrix for the noise. The first term in the exponent, which depends only on the GWs, is identified as $\snr^2/2$, while the second term is a random variable with mean zero and variance (over noise realizations) equal to $\snr^2$. This can be proved, e.g., by considering that Gaussian noise with covariance $C$ can be written as $\sqrt{C} \, \bar{n}$, with $\sqrt{C} \sqrt{C}^T = C$ the Cholesky decomposition of $C$, and with $\bar{n}$ a vector of uncorrelated, zero-mean/unit-variance Gaussian variables. Then
\begin{eqnarray*}
&& \langle (n_i (C^{-1})^{ij} g_j) (n_l (C^{-1})^{lm} g_m) \rangle \\ &&\quad =  
(C^{-1})^{ij} g_j \sqrt{C}_i^k \langle \bar{n}_k \bar{n}_p \rangle
\sqrt{C}_l^p (C^{-1})^{lm} g_m \\ &&\quad =
(C^{-1})^{ij} C^{il} (C^{-1})^{lm} g_j g_m \\ &&\quad = (C^{-1})^{jm} g_j g_m~.
\end{eqnarray*}

Equation \eqref{eq:snr} follows immediately under the (strong) assumption that noise is uncorrelated and homogenous among pulsars and epochs, so that it can be represented by $(C^{-1})^{ij} = \delta^{ij} / \delta t^2_\rms$. We are assuming also that the sampling of pulsars and epochs in the dataset is sufficiently broad and non-pathological that $\sum_i g^2_i \simeq M N \langle \delta t^2_\gw \rangle$; that is, that the sampling can effectively perform an average over time and pulsar sky position. If the noise is uncorrelated (i.e., \emph{white}), but not homogeneous, Eq.\ \eqref{eq:snr} still stands, provided that $\delta t^2_\rms$ can be taken to represent a suitable \emph{averaged} noise.

Under these assumptions, Eq.\ \eqref{eq:snr} is remarkable in that the actual \emph{form} of the signal to be detected appears only through its variance $\langle \delta t^2_\gw \rangle$, and that the structure of observations appears only through their overall number $M \times N$ and rms noise $\delta t^2_\rms$. By contrast, one may have imagined that detecting (say) quasi-sinusoidal signals of high frequency $f_\gw$ would require rapid-cadence observations spaced by $\Delta t \lesssim 1/f_\gw$, according to the Nyquist theorem. However, that theorem is a statement about the \emph{reconstruction} of the whole of a function on the basis of a set of regularly spaced samples, but it does not apply to our case---computing the likelihood that a signal of known shape is present in the data \cite{2001AIPC..567....1B}. In effect, we are checking that the measured data are consistent with our postulated signal: for uncorrelated noise, it does not matter \emph{when} we check, but only \emph{how many times} we do it.

\subsection{Relaxing the assumption of white noise}\label{sec:relax}

There are two important considerations that challenge our assumption of white, uncorrelated noise.

The first is that the residuals include a stochastic contribution due to the over-fitting of noise (and possible GWs) at the time of deriving the timing model. We discuss this further in Sec.\ \ref{sec:degen}, where we show empirically that the detection of quasi-sinusoidal signals at most frequencies would not be affected.
From a formal standpoint, van Haasteren and colleagues
\cite{2009MNRAS.395.1005V} show that it possible to \emph{marginalize}
the likelihood over timing-model parameter errors $\delta \xi$ by
replacing the inverse covariance in Eq.\ \eqref{eq:likes} with $C^{-1}
- C^{-1} M (M^T C^{-1} M)^{-1} M^T C^{-1}$, where $M$ is the
\emph{design matrix} for the timing model fit, so that the extra contribution to the residuals has the form $M \delta \xi$ 
(A similar strategy of ``projecting out'' parameter errors was 
employed earlier by Cutler and Harms~\cite{2006PhRvD..73d2001C}, in the context of removing residual noise from slightly incorrect GW foreground subtraction).
For uncorrelated noise, Eq.\ \eqref{eq:snr} is modified only by restricting the computation of $\langle \delta t^2_\gw \rangle$ to the GW components that are not absorbed away by the timing model (and this is indeed what we investigate in Sec.\ \ref{sec:degen}).

The second important consideration (and for which the GW frequency
\emph{does} matter) is the impact of correlated noise. The physically
interesting case here is that of long-term correlations, which
generate \emph{red} noise that is stronger at low frequencies. To
understand the impact of red noise, we study a toy model in which the $N$ observations are organized in $P$ ``clumps'' of $Q$ TOAs taken at nearby times (with $N = P \times Q$), and where noise consists of two components: uncorrelated noise with variance $\sigma^2$ and noise with variance $\kappa^2$ that is completely correlated within clumps, and completely uncorrelated between clumps. (We use $\kappa$ since $\kappa \acute{o} \kappa \kappa \iota\nu o\zeta$ is Greek for ``red.'')
 We consider a single pulsar, although the generalization to more is trivial.

The resulting $C$ has the structure
\begin{equation}
C = \sigma^2 I + \kappa^2 \sum_{i=1}^{P} O_i,
\end{equation}
where each $O_i$ is a matrix that has ones for every component corresponding to a combination of samples in the same burst, and zeros everywhere else. Each $O_i$ can also be written as $u_i u_i^T$, where $u_i$ is a vector that has ones for the components in clump $i$, and zeros everywhere else. From the block structure of $C$ and the Woodbury lemma \cite{hager1989}, it follows that
\begin{equation}
C^{-1} = \sigma^{-2} I - \frac{\sigma^{-2}}{P + \kappa^{-2}/\sigma^{-2}} \sum_{i=1}^{P} O_i.
\end{equation}
If the characteristic frequency of the GW signal is ``slower'' than the timescale of a clump (i.e., the time over which the $Q$ samples in a clump are collected), then the sum $\sum_i g_i^T O_i g_i \simeq P Q^2 \langle \delta t_\gw^2 \rangle$, because the same value of $g$ is being summed over and over in each burst.
%After some algebra, it follows that
It follows that
\begin{equation}
\label{eq:snrmod}
\snr^2 = \frac{\langle \delta t_\gw^2 \rangle PQ}{\sigma^2 + Q \kappa^2}
= \frac{\langle \delta t_\gw^2 \rangle P}{\sigma^2/Q + \kappa^2};
\end{equation}
that is, the repeated observations in each clump average out the
uncorrelated component of noise (as $\propto 1/\sqrt{Q}$), but not its
correlated part. Increasing the number of observations in a clump provides diminishing returns as $\sigma^2/Q \rightarrow \kappa^2$.

Let us follow the other branch of our derivation: if the
characteristic frequency of the GW signal is ``faster'' than the
timescale of the clumps, then, barring special coincidences, $\sum_i g_i^T O_i g_i \simeq P Q \langle \delta t_\gw^2 \rangle$, and $\snr^2$ reduces (modulo an $O[1/Q]$ correction) to the general expression \eqref{eq:snr}, with $N = PQ$.

\subsection{Noise characteristics inferred from observational data}\label{sec:TN}

In this section, we consider the characteristics of noise for real
pulsars.  Namely, to what extent is our analysis applicable to timing
residuals from actual PTAs?

For the \emph{radiometer} noise due to thermal effects in the
receiving system, the assumption of no correlations (i.e., ``white'') is well justified: for observations over a radio frequency bandwidth $\Delta\nu$, the correlation timescale is $(\Delta\nu)^{-1}$, so this noise contribution is effectively uncorrelated in time.
%Cordes and Shannon \cite{2010arXiv1010.3785C} show that typical radiometer noise has $\sigma \sim 1\,\mu\mathrm{s}$.
Further, from thermodynamic considerations, the assumption of
Gaussianity is also well justified.

Pulsars can show correlated, red spectrum fluctuations in their TOAs,
and 
Cordes and Shannon \cite{2010arXiv1010.3785C} present a summary of
various effects, ranging from intrinsic spin fluctuations to magnetospheric and propagation effects; see also \cite{2010ApJ...725.1607S}. These effects have spectral densities $\propto f^{-x}$, with $x$ typically $> 1$ and in some cases $> 4$.
On timescales $\sim$ 5 years ($f \sim 10^{-8.2}\,\mathrm{Hz}$), the residuals appear to be dominated by white components (\citep{2009MNRAS.400..951V,2013ApJ...762...94D}; see also Figs.\ 10 and 11 of \citep{2013PASA...30...17M} for a visual representation of noise effects in PPTA pulsars).
Even if $\sigma^2 \approx \kappa^2$ at frequencies $\sim
10^{-8.2}\,\mathrm{Hz}$, at higher frequencies ($\gtrsim
10^{-7}\,\mathrm{Hz}$), the variance from white processes will exceed
that of any red processes with relatively shallow spectra ($x \approx
1$) by a factor of approximately 15; for red processes with steeper
spectra ($x \approx 4$), the ratio will be even larger.
\red{A recent global experiment in which several telescopes observed PSR J1713+0747 over the course of 24 hours presents an opportunity to test the extent to which white-noise processes dominate potential red-noise processes.}

%\blue{
In our toy model, the red-noise component of the variance is amplified by the clump multiplicity $Q$ [Eq.\ \eqref{eq:snrmod}]. For more general observation schemes and red-noise processes, we may think of the number of clumps $P$ as $T_\mathrm{obs} / T_\mathrm{red}$, where $T_\mathrm{obs}$ is the total duration of observation, and $T_\mathrm{red}$ is the correlation timescale of the most significant red-noise process; then $Q \simeq N (T_\mathrm{red} / T_\mathrm{obs})$. For GW signals with frequency $\lesssim 1/T_\mathrm{red}$, our toy model would then suggest that
\begin{equation}
\label{eq:snrmodtwo}
\snr^2 = \frac{\langle \delta t_\gw^2 \rangle}{\sigma^2/N + \kappa^2 (T_\mathrm{red}/T_\mathrm{obs})};
\end{equation}
that is, the $1/\sqrt{N}$ averaging of noise becomes limited by red
noise once $N \sim (\sigma^2/\kappa^2) \times
(T_\mathrm{obs}/T_\mathrm{red})$---an interesting scaling in its own
right.
%This scaling implies that red-noise processes are unimportant,
%provided that the number of data acquired per clump~$Q$ is smaller
%than the ratio~$\sigma^2/\kappa^2$ (since $P \sim T_\mathrm{obs} /
%T_\mathrm{red}$).
For GW signals with frequencies $\gtrsim 1/T_\mathrm{red}$, the simpler scaling \eqref{eq:snr} applies.
%}

In the remainder of this paper, we neglect the effects of red noise in the scaling of \hbox{SNR} and assume the expression of Eq.\ \eqref{eq:snr}.  Our assumption is correct because one or more of the following circumstances will be true (or true \emph{enough}) in practice:
\begin{itemize}
\item The characteristic GW frequency of interest will be greater than
$1/T_\mathrm{red}$ for the most significant red-noise component.
\item For a majority of the pulsars in the \hbox{PTA}, the white-noise
variance will exceed that of the most dominant red-noise process for
the time scales of interest.
%SBS: That is, of course, only for pulsars with DM correction (most have this, however, or at least will have this in the future as people have begun to realize how much DM variation contributes to red noise variations...... so I agree with the written statement.)
%
\item The number of observations will not saturate the
averaging of white noise with respect to sub-dominant red noise (i.e.,
in the ``clump'' picture, $\sigma^2/Q > \kappa^2$).
\end{itemize}

\section{Brief review of prospects for PTA searches of supermassive--black-hole binaries and cosmic strings}
\label{sec:source-review}

Here we collect a few salient points concerning PTA searches for SMBHBs and 
cosmic strings, mostly to provide points of comparison with the hypothetical sources we consider in the next sections.  We refer the reader to the literature cited below for more details.

\subsection{The detectability of GWs from supermassive black hole binaries}

When two galaxies merge, the SMBHs at their centers are brought together by tidal friction
from the surrounding stars and gas.  It seems likely that their
separation eventually shrinks to the point at which gravitational radiation emission dominates the inspiral, and the two 
SMBHs eventually coalesce \cite{1980Natur.287..307B}.
The GWs from all in-spiraling SMBHBs in the observable Universe contribute to a stochastic background of GWs
with characteristic amplitude $h_c \sim h_\rms \sqrt{f}$ given by
\be \label{hc_smbh}
h_c \approx A (f/f_0)^{-\beta}
\ee
in the PTA band, where  $\beta \approx {2/3}$ and $A$ is predicted to be in the range $5 \times 10^{-16}\mbox{--}5 \times 10^{-15}$ for $f_0 = 10^{-8}$ Hz \cite{2003ApJ...583..616J,2003ApJ...590..691W,2012arXiv1211.4590M,2013MNRAS.433L...1S}. Depending on the actual $A$, the first PTA detection of GWs is expected between 2016 and 2020 \cite{2013arXiv1305.3196S}.

The background is expected to be dominated by binaries with chirp masses $M_c \equiv (m_1 m_2)^{3/5} (m_1 + m_2)^{-1/5} \sim 10^8 M_\odot$ at $z \alt 2$. At frequencies above $f \approx 10^{-8}$ Hz, sources are sparse enough that the central limit theorem does not apply, so the distribution is significantly non-Gaussian and a few brightest
sources would appear above the background. Thus, the first PTA discovery could either be 
an individual strong (and possibly nearby) source, or the full background.

\subsection{The detectability of GWs from cosmic strings}
There are several mechanisms by which an observable network of cosmic (super)strings could have formed in the early Universe~\cite{2007arXiv0707.0888P}.
% Basically, string formation arises from the breaking of some U(1) symmetry (either global or local) as the Universe expands and cools.  In the 1980s and 1990s, interest was primarily in cosmic strings arising from grand unified theories~\cite{VilenkinShellard}, but in recent years several string-theory-inspired inflationary models have also been shown to populate the Universe with a network of cosmic-scale strings~\cite{SarangiTye,Polchinski04},  
Simulations have shown that string networks rapidly approach an attractor: the distribution of straight strings and loops in a Hubble volume becomes independent of initial conditions.  The network properties {\it do} depend on two fundamental parameters: the string tension $\mu$ and the string reconnection probability $p$. 
The size of string loops at their birth should in principle be derivable from $\mu$ and $p$, but the studies are difficult and  different simulations have produced very different answers.  Therefore most astrophysical analyses today assume that the size of loops at their birth can be parametrized as $\alpha \, H^{-1}(z)$, where $H^{-1}(z)$
is the Hubble scale when the loop is ``born,'' and where $\alpha$ is treated as a third unknown parameter.  We refer the reader to ~\cite{1997stgr.proc....3A,2007arXiv0707.0888P} for nice reviews.
To make matters more complicated, Polchinski has argued that the distribution of loop size at birth is actually bimodal, with both relatively large and small loops being produced at the same epoch~\cite{2008PhRvD..77l3528D}.    
Regarding the string tension $\mu$, physically motivated values range over at least six orders of magnitude: %which could reasonably have string tensions in the range 
$10^{-12} \lesssim \mu \lesssim 10^{-6}$. 
%For instance, brane-inflation models can naturally lead to the breaking of U(1) symmetries at %the end of inflation, leading to the formation of both long fundamental strings and $D(k+1)$-%branes that wrap around $k$ compact dimensions and extend in one of Nature's %three large spatial dimensions. These long strings can be stable on cosmological timescales %(depending on the exact model) and could reasonably have string tensions in the range %$10^{-12} \lesssim \mu \lesssim 10^{-6}$.  

%For nice reviews we refer to \cite{Allen_review,VilenkinShellardPolchinski04_review} .
%
%huge range of scales makes this difficult to simulate, simulations, and today the typical loop size at birth (as a fraction of the Hubble scale) is still uncertain by many orders of magnitude. Polchinski has pointed out that the distribution could be bi-modal,
% It could well be bi-modal,
%with both relatively large and small loops being produced at the same epoch. (ref to Polchinkski)
%
%For nice reviews we refer to \cite{Allen_review,VilenkinShellardPolchinski04_review} .
%
% We refer the reader to Allen~\cite{Allen_review} for a brief, pedagogical introduction to string %networks, and to Vilenkin and Shellard \cite{VilenkinShellard} for a more comprehensive review.

Once formed, string loops oscillate and therefore lose energy and shrink due to GW emission.  These waves form a stochastic GW background.  In addition to this approximately Gaussian background, the cusps and kinks that form on the string loops emit highly beamed GW bursts~\cite{2001PhRvD..64f4008D,2005PhRvD..71f3510D} Depending on the string parameters, PTAs could discover the stochastic background, the individual bursts, both or neither. 
The current limit on  $\Omega_\mathrm{GW}(f)$ from pulsar timing is $\Omega_\mathrm{GW}(f\sim1\,{\rm yr}^{-1}) \lesssim  1\times 10^{-8}$~\cite{2011MNRAS.414.3117V}, corresponding to a limit on the string tension of $G\mu\leq4.0\times10^{-9}$.
%!!!\red{Need to change this to shannon et al. reference when it's out.}

 %The spectrum of this GW background radiation is calculated to be roughly flat over many orders %of magnitude in frequency, including the frequency bands where current ground-based GW %interferometers (like LIGO and Virgo) and planned space-based GW interferometers (like LISA) %are sensitive. 
%It is conventional to express the energy density $\rho_{GW}$ of GWs in terms of 
%\be 
%\Omega_\mathrm{GW}(f) \equiv \frac{1}{\rho_c}\frac{d  \rho_\mathrm{GW}}{ d \, {\rm ln}\, f} \, ,
%\ee
%where $\rho_c$ is the Universe's closure density.
%By comparison, the limit from first-generation ground-based interferometers is %$\Omega_\mathrm{GW}(f\sim 100\,\mathrm{Hz}) < 6.9\times 10^{-6}$ \cite{2009Nature}.
%For comparison, the Advanced LIGO detectors should be capable of detecting a stochastic %background with $\Omega_\mathrm{GW}(f\sim %40 \, \mathrm{Hz}) \gtrsim %10^{-9}$~\cite{2009Nature}, \red{this is 3 yrs old and so should be updated}.

\begin{figure*}
\centering
\includegraphics[width=0.7\textwidth]{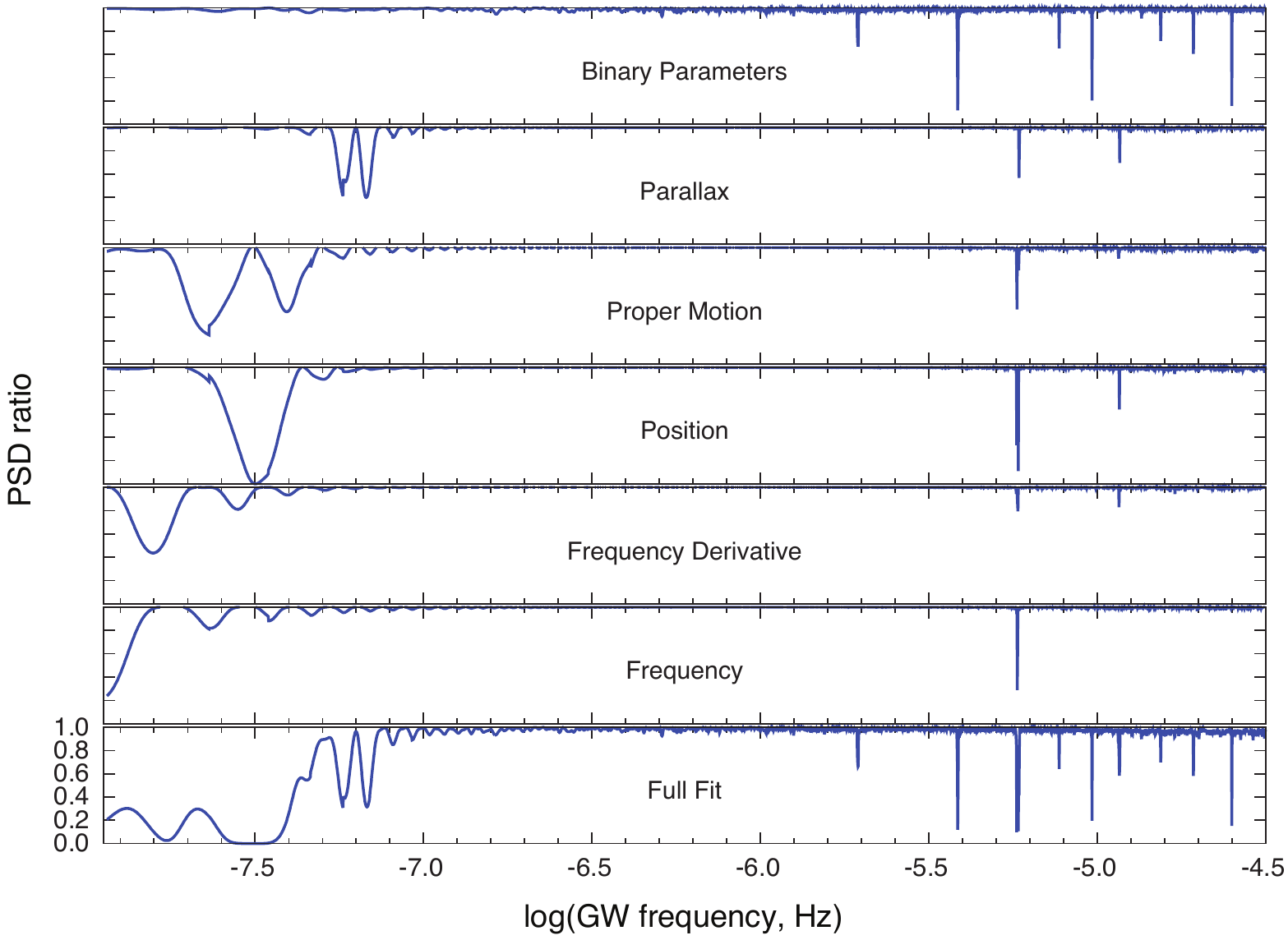}
\caption{GW power absorbed by fitting for various pulsar parameters as a function of GW frequency, for pulsar J0613$-$0200. ``PSD ratio'' refers to the pre-fit power spectral density value for the given frequency, divided by its post-fit value. All simulated GWs were sinusoids at the given GW frequency. For each panel, only the indicated parameters were used for fitting, while the other parameters were held fixed at the values given in \cite{2013PASA...30...17M}.  At high frequencies, only narrow features are evident (mostly due to fitting of the pulsar's binary motions), but low-frequency GW signals are significantly absorbed by standard fitting parameters.
\label{fig:frange}}
\end{figure*}
\red{Curt will probably just leave out next paragraph, but not 100\% sure yet.  If included, he needs to check that his factor 10 agrees with what's in Hogan's paper:
Note that the above limits assumed that cosmic strings are uniformly distributed within the 
Universe. However Hogan and dePies pointed out that for $\mu \alt 10^{-12}$, the recoil of string loops due to GW emission is sufficiently small that their center-of-mass velocity 
drops to the point at which they cluster around galactic halos. 
If so, at the Earth the GW signal from strings is dominated by loops within our own Galactic halo.  As shown more generally in Sec.\ \ref{galactic}, the result is to increase the estimate of the SNR from strings by a factor $\sim 10$, compared to the estimate based on a uniform distribution (of course, for the same $\mu$, $\alpha$, and $\Omega_{\gw}$.  }

\subsection{Current constraints on $\Omega_\mathrm{GW}(f)$}

As mentioned, the current limit on  $\Omega_\mathrm{GW}(f)$ from pulsar timing is $\Omega_\mathrm{GW}(f\sim1\,{\rm yr}^{-1}) \lesssim  1\times 10^{-8}$~\cite{2011MNRAS.414.3117V}.
%!!!\red{[Must update with Shannon once it's out.]}
By comparison, the limit from first-generation ground-based interferometers is $\Omega_\mathrm{GW}(f\sim 100\,\mathrm{Hz}) < 6.9\times 10^{-6}$ \cite{2009Natur.460..990A}.

From Big Bang nucleosynthesis, we know also that any GW stochastic background that existed already when the Universe was three minutes old satisfies $\Omega_\mathrm{GW} < 1.5 \times 10^{-5}$ today~\cite{2008PhRvD..78d3531B}.
Combined measurements of CMB angular power spectra (which are sensitive to lensing by a stochastic GW background) with matter power spectra also yield $\Omega_\mathrm{GW} \alt 10^{-5}$ today, but this method is sensitive to any GWs produced before recombination at $z \approx 1100$~\cite{2006PhRvL..97b1301S}.
For GWs generated in the low-z universe, combining results from Planck, WMAP, SDSS, and $H_0$ measurements gives the limit $\Omega_\mathrm{GW} \alt 6 \times 10^{-3}$~\cite{2013arXiv1307.0615A}.
% Here's Curt's quick summary of how he got 6e-3:  It would take sum_neutrino_masses = 94 ev to make Omega_{neutrino) h^2 = 1. This ref gives 95% conf bound on sum of 0.27eV, using all the above expts.  So Om < 0.27/(94 h^2) ~ 6e-3.
\red{[Curt: could add for completeness CMB bound on extreme low frequencies: $\Omega_{\gw} < 10^{-14}$ for $f< 10^{-16}$ Hz.]}
% Rotti and Souradeep~\cite{2012PhRvL.109v1301R} say CMB also gives constraints on GWs sourced between z=1100 and z=1, but only for very low f: very roughly $\Omega_{\gw} \alt 10^{-10} --10^{-6}$ for $f$ in range $10^{-17} --10^{-15}\,$Hz.  Hard to say better from the figure: ULs(source at z) fill out a parallelogram in $\Omega-f$ space.  Also there's some LIGO limit.

\section{Spectral absorption effects from pulsar timing-model fitting}
\label{sec:degen}

The best knowledge of pulsar parameters comes from the iterative observation and refinement of a timing model, which predicts the times of arrival of all the pulses as a function of all relevant parameters, such as the period and period derivatives of the pulsar's intrinsic spin; the position, proper motion, and parallax of the pulsar; and possibly parameters that describe the motion of the pulsar in a binary system.
Depending on the cadence and total time of observation, and on the shape and duration of the GWs, the effects of the GWs on pulse arrival times may correlate with the effects of changing the pulsar parameters, so the GW power may be partly on entirely absorbed by the parameter-fitting process (see, e.,g., the study of the effect of a GW background on pulsar timing parameter estimation \cite{2011MNRAS.417.2318E}).

%\red{[Curt: we want to make point that any signal is sum of sinusoids, so if a sinusoid can't be hidden by param errors, then no signal can be hidden.  Basic picture from our sys-error paper. Michele: but clearly some signals can be hidden (although sinusoids may not be the best description for them), so here we're showing that some sinusoids \emph{of interest} are not affected.]}
%SBS: I changed the next sentence to be more generic.

As a specific study of this effect, here we investigate the absorption of sinusoidal GWs to demonstrate frequency-dependent signal loss to pulsar parameter fitting. To do this, we use the {\sc Tempo2} software suite \citep{2006MNRAS.369..655H} to simulate a set of timing residuals for pulsar J0613-0200 \cite{2013PASA...30...17M}, as observed with the Parkes observatory. We generate one TOA every other day for $T_\mathrm{obs} =$ 1,000 days, at a random time compatible with the pulsar being visible from the observatory, and we add a white-noise component with rms amplitude of 100 ns.
Into these simulated residuals we inject sinusoidal GWs from a circular SMBHB binary located at $\mathrm{RA} = \mathrm{Dec} = 0$, varying the GW frequency $f$ between $10^{-7.95}$ and $10^{-4.5}$ Hz (corresponding to GW periods of $\sim$ 1,000 days to eight hours), and setting $h_{+} = h_{\times} = 10^{-3} f$, so that the SNR is fixed.

For each GW frequency, we measure the power spectral density of the relevant frequency component before the timing-model fit and after seven different types of fit: a fit against the full set of parameters and individual fits for pulsar frequency, frequency derivative, position, proper motion, parallax, and binary period. Figure~\ref{fig:frange} shows the ratio of the power spectral densities in each case, as a function of the source GW frequency. In effect, we are showing the \emph{absorption spectrum} of sinusoidal GWs, as filtered by the timing-model fit.
Above $10^{-7}$\,Hz, $\gtrsim 95\%$ of the signal is preserved even in a full parameter fit, with only narrow absorption features. It is clear that most of these features are specific to this pulsar's binary orbit (and its harmonics), and would not appear at the same frequencies for other pulsars in a PTA.

However, absorption features originating from non-binary parameters will occur in all pulsars. Specifically, absorption at $f = 1 / \mathrm{year}$ (corresponding to pulsar position/proper motion) and $1 / 6$ months (corresponding to parallax) can result in up to 100\% loss of the GW signal. Similarly, as the GW period approaches the total duration of pulsar observations, fitting the pulsar spin frequency and frequency derivative results in significant signal absorption. The sensitivity to GWs at these lower frequencies would be better in a longer data set (see, e.g., the low-frequency sensitivity curves in \cite{2010MNRAS.407..669Y}).

At high frequencies, only two narrow absorption features may be common across a PTA: these correspond to the observing cadence (here at $1/(2\,{\rm days})=10^{-5.238}$ Hz), and to the sidereal day (at $1/(23.934 \, \mathrm{hr})\simeq10^{-4.935}$ Hz). The former can be avoided with higher-cadence or irregular observations, but the latter reflects the limitations of using a single observatory, which can only observe a source while it is above the horizon. In our simulation we have chosen random observation times within the window of coverage, but more structured observing cycles can engender even deeper features. By contrast, this feature can be avoided for a polar target that never sets.

To summarize, our example study suggests that for a majority of PTAs a high-frequency GW signal will be well preserved through the standard timing-model fitting process, save for narrow features at roughly the observing cadence and the sidereal day. GWs at frequencies close to either (1\,year)$^{-1}$) or (6\,months)$^{-1}$ will be significantly impacted, as will GWs with periods approaching the longest-duration pulsar observations. 
%\red{MV: question for Sarah---is there any precedent to this kind of study in the literature?}
%SBS: I added a reference earlier to Ellis et al (2010); not the same kind of study or goal, but analysis was close enough.

\section{Discovery space for sources in the low-redshift Universe, $z \alt \emph{O}(1)$}
\label{sec:z1}

In this section we begin to characterize the PTA discovery space for the case of sources in the low-redshift Universe, by which we mean
$z \alt \emph{O}(1)$.
We imagine that there is \emph{some} heretofore undiscovered GW source, and we ask what it would take for it to be detectable via pulsar timing. We consider separately the case of \emph{modeled signals} (for sources already conceived by theorists, so that a parameterized waveform model can be used in a matched-filtering search), the case of \emph{unmodeled bursts}, and the case of the \emph{gravitational memory} effect from modeled sources.

 We will assume that the GW sources are distributed isotropically and that we do not occupy a preferred location in space \emph{and} time with respect to them---that is, we assume that the Earth is not improbably close (spatially) to one of the sources, and that the sources have been emitting GWs for a significant fraction of the last $10^{10}$ years.

We parametrize our projections in terms of the energy density $\Omega_\mathrm{GW}$. 
Because we consider sources in the low-redshift Universe, in what follows we ignore redshift effects.
Nevertheless our results at $z \sim 1$ match on nicely to our results for high-z in Section VII.
  
\subsection{Discovery space for modeled GW signals in the low-redshift Universe}
\label{sec:gwsig}

As we established in Sec.\ \ref{sec:determ}, the SNR modeled GW signals as observed by a PTA is
\newcommand{\sig}{\mathrm{sig}}
\newcommand{\obs}{\mathrm{obs}}
\begin{equation}
\label{eq:snr2}
\begin{aligned}
\snr^2 &= \frac{\langle \delta t_\gw^2 \rangle}{\delta t_\rms^2} M N
\\ &= \frac{\langle \delta t_\gw^2 \rangle}{\delta t_\rms^2} M p \min \{T_\sig,T_\obs\},
\end{aligned}
\end{equation}
where $\langle \delta t_\gw^2 \rangle$ and $\delta t_\rms^2$ are the mean-square-averaged timing residuals due to GWs and measurement/pulsar noise; $M$ is the number of pulsars in the array; $N$ is the number of times each pulsar is observed, which we rewrite in terms of the cadence of observation $p$ (e.g., 1/day), the total duration of observation $T_\obs$ (e.g., 3 years), and the typical duration of the GW signal $T_\sig$. 

For a sinusoidal GW signal of frequency $f$ and rms amplitude at Earth $h = \sqrt{h_+^2 + h_\times^2}$, the root-mean-square timing residual averages\footnote{To derive Eq.\ \eqref{eq:dt} we compute the Estabrook--Wahquist \cite{1975GReGr...6..439E} fractional Doppler response (for the pulsar ``Earth term'' alone) to a sinusoidal GWs given by $h_+(t) + i h_\times(t) = (h/\sqrt{2}) \exp 2 \pi i f t$, take the antiderivative to obtain the corresponding pulse-time delay, square and average over time, sky position, and polarization angle. \red{[MV: this specification of $h_+(t)$ and $h_\times(t)$ may too restrictive; my calculation should be valid whenever $h_+$ and $h_\times$ are uncorrelated and have the same amplitude $= h/\sqrt{2}$.}} to
\begin{equation}
\label{eq:dt}
\bar{\delta} t_\gw \equiv
\sqrt{\langle \delta t_\gw^2 \rangle} =
\frac{1}{4\sqrt{3}\pi} \frac{h}{f} \simeq
\frac{1}{20} \frac{h}{f}.
\end{equation}
Furthermore, the average rate at which the sources radiates energy in GWs is $\dot{E} = (\pi^2/2) h^2 f^2 d^2 \simeq 5 h^2 f^2 d^2$ \cite[Eq.\ (1.160)]{maggiore2008}, where $d$ is the distance to the source, and $G = c = 1$ (as we will set throughout). The GW energy density from source of this kind is
\begin{equation}
\Omega_\gw \rho_0 \simeq (\dot{E} T_\sig)(R_4 \tau_0),
\end{equation}
where $R_4$ is the spacetime rate--density of sources, and 
%$\tau_0 \sim 1.4 \times 10^{10}$ yr is the current age of the Universe.
$\tau_0 \sim 10^{10}$ yr is the current age of the Universe.
Approximating the closure density $\rho_0 = 3 H^2/{8 \pi}$ as $\tau_0^{-2}/10$ (since $\tau_0 \simeq H^{-1}$) and rewriting $R_4 \equiv (V_R T_R)^{-1}$ in terms of a fiducial volume $V_R$ and the total event rate $T_R$ in that volume, we can re-express the expected GW-induced timing residual as
\begin{equation}
\label{eq:dtOM}
\bar{\delta} t_\gw \simeq \frac{1}{150} f^{-2}\, d^{-1}\, \bigg(\frac{\Omega_\gw V_R T_R}{\tau_0^3\, T_\sig}\bigg)^{1/2} \, .
\end{equation}

We estimate the distance to the closest source that would be observed over time $T_\obs$ by setting
\begin{equation}
\frac{4}{3} \pi \, d^3 \max \{T_\obs,T_\sig\} R_4 = 1
\end{equation}
(where the maximum accounts for the persistence of multiple emitting sources if $T_\sig > T_\obs$), whence
\begin{equation}
\label{eq:dnear}
d_\mathrm{near} \simeq 
\left[
\frac{3}{4\pi} \frac{V_R T_R}{T_\sig} \min\{1,T_\sig/T_\obs\}
\right]^{1/3}.
\end{equation}
Folding together all the results of this section, we obtain the corresponding, largest SNR that would be observed as
\begin{equation}
\label{eq:snrnear}
\begin{aligned}
\snr^2_\mathrm{near}
& \simeq
10^{-4} \frac{\Omega_\gw}{f^4 \tau_0^3}\, \frac{M p\, T_\obs}{\delta t^{2}_\rms}
\bigg[ \frac{V_R\, T_R}{T_\sig}  {\rm min}\{1,\frac{T_\sig}{T_\obs}\}\bigg]^{1/3} \\
& \simeq
2 \times 10^{-4} \frac{\Omega_\gw}{f^4 \tau_0^3}\, \frac{M p\, T_\obs}{\delta t^{2}_\rms} d_\mathrm{near}.
\end{aligned}
\end{equation}
We would now like to determine how large an $\snr_\mathrm{near}$ we could expect for a given $\Omega_\gw$, and for given observational parameters $M$, $p$, $T_\obs$, and $\delta t^2_\rms$. This amounts to maximizing $\snr_\mathrm{near}$ with respect to the GW-source parameters $V_R$, $T_R$, and $T_\sig$; since these appear together in $d_\mathrm{near}$, we obtain the largest possible $\snr_\mathrm{near}$ by setting $d_\mathrm{near} = \tau_0$, the Hubble distance. We dare not place the GW source farther, since we are considering the ``local'' Universe and neglecting redshift effects.

Note that the scaling $\snr^2_\mathrm{near} \propto d_\mathrm{near}$ of Eq.\ \eqref{eq:snrnear} seems counterintuitive, since we would naively think of the strongest sources as the closest. However, while the squared GW strain $h^2$ at the Earth scales as $1/d^2$, it also scales with the total energy $\Delta E$ that is emitted by each source, and that is ``available'' to each source given a fixed $\Omega_\gw$; this $\Delta E$ increases with decreasing source density, and is proportional to $d_\mathrm{near}^3$. This surprising intermediate result was already shown in ZT82 \cite{Zimmermann:1982wi}.

We can now plug in fiducial values for the observational parameters (as well as $\tau_0 = 3 \times 10^{17}$ s), arriving at
\begin{equation}
\label{eq:fid1}
\begin{aligned}
\max\{\snr \}& \lesssim 10 \bigg(\frac{f}{10^{-7} \, {\rm Hz}}\bigg)^{\!\!-2}
\bigg[\frac{\Omega_\gw}{10^{-5} }\bigg]^{1/2} \times \mathrm{obs.}   \\
 & \lesssim 0.03 \bigg(\frac{f}{10^{-5} \, {\rm Hz}}\bigg)^{\!\!-2}
\bigg[\frac{\Omega_\gw}{10^{-2}}\bigg]^{1/2} \times \mathrm{obs.}, 
\end{aligned}
\end{equation}
where
\begin{equation}
\mathrm{obs.} = \bigg[\frac{\delta t_\rms}{10^{-7} \, \mathrm{s}}\bigg]^{-1}
\bigg[\frac{M \,  p\, T_\obs }{10^{4}}\bigg]^{1/2}.
\end{equation}
While we derived these constraints for the case of small $z$, we shall see below that they become even stronger for high-$z$ sources.

The fiducial values for $f$ and $\Omega_\gw$ in the second row of Eq.\ \eqref{eq:fid1} are motivated by our original question, whether PTA searches should be extended to frequencies as high as $\sim 10^{-5}$ Hz. The current upper limit (from structure formation) on the energy density of hot dark matter is $\Omega_\mathrm{HDM} \alt 1.5 \times 10^{-2}$ (at $95\%$ confidence) \red{[need ref]}; this limit applies also to $\Omega_\gw$. Our conclusion is that a PTA detection of GWs at frequencies above $\sim 3 \times 10^{-5}$\,Hz should be considered very unlikely on fundamental grounds.

\subsection{Discovery space for unmodeled GW bursts in the low-redshift Universe}

Quite simply, a burst is a signal with $T_\sig \sim 1/f$. Since it contains only $\sim 1$ cycles, its instantaneous SNR (i.e., GW amplitude over rms noise) is the same as its matched-filtering SNR, up to a factor of order one (after the data has been filtered to remove the noise that is outside the band of interest). Now, whatever the $T_\sig$, we can still adjust $R_4$ so that $d_\mathrm{near}$, as defined in Eq.\ \eqref{eq:dnear}, is equal to $\tau_0$.

For instance, if $T_\sig \lesssim T_\obs$ and $T_\obs = 10^8$ s, this requires one burst every $10^8$ s within a Hubble volume. So for this rate, the instantaneous SNR is the same as given in Eq.\ \eqref{eq:fid1} for modeled signals. This seems promising, because since bursts require no model for their detection, they could potentially reveal phenomena that nobody ever thought of. At the same time, their detection would require the utmost care in excluding instrumental and astrophysical artifacts.

\subsection{Discovery space for GW memory in the low-redshift Universe}

GWs with memory (for a recent review see \cite{2010CQGra..27h4036F}) cause a permanent deformation -- a ``memory'' of the passage of the waves -- in the configuration of an idealized GW detector. They are emitted by systems with unbound components (linear memory), and by \emph{generic} GW sources because of the contribution of the energy--momentum of their ``standard'' GWs to the changing radiative moments of the source (nonlinear memory). 
Several authors have discussed the detectability of the GW memory effect by PTAs for known source types, especially merging massive--black-hole binaries~\cite{2009MNRAS.400L..38S,2010MNRAS.401.2372V,2010MNRAS.402..417P,2012ApJ...752...54C}. Here we consider the effect from the point of view of the PTA discovery space, and again we ask in which region of parameter space PTAs could discover previously unimagined sources by way of their GW memory.

For a source at distance $d$ from Earth, which emits a total energy of $\Delta E$ in GWs, the amplitude of the memory effect is \cite{2010MNRAS.401.2372V}
\newcommand{\mem}{\mathrm{mem}}
\begin{equation}
\label{eq:mem1}
h_\mem \sim \frac{\alpha}{\sqrt{6}} \frac{\Delta E}{d},
\end{equation}
where $\alpha < 1$ is a factor determined by the asphericity of the energy outflow (more precisely, from its quadrupolar part).

In addition to the general assumptions we made in Sec.\ \ref{sec:z1}, we will postulate that most of the GW energy from any one source is emitted on a timescale $T_\sig \ll T_\obs$. 
Then we can approximate the ``turn on'' of the memory effect as a step function, and the effect on any pulsar is to create a timing residual that grows linearly in time:
\begin{equation}
\label{eq:mem2}
\delta t_\gw \sim \theta(t-t_0) h_\mem t,
\end{equation}
where the memory passes over the Earth at time $t_0$.

In any single pulsar, a linear-in-time residual can be interpreted simply as a glitch causing an instantaneous change in the pulsar frequency. However, all pulsars in the PTA would show such apparent glitches at the same time, with relative amplitudes following a simple pattern on the sky~\cite{2010MNRAS.401.2372V} determined by four parameters (the sky-location angles and two amplitudes that specify the transverse--trace-free part of the metric), so in principle the detection problem is well posed.
The corresponding PTA SNR is \cite{2010MNRAS.401.2372V}
\begin{equation}
\label{eq:mem-snr}
\snr_\mem \sim \frac{1}{20}  \frac{h_\mem \, T_\obs}{\delta t_\rms} (M p\, T_\obs)^{1/2} \, ,
\end{equation}
where the factor $1/20$ accounts for the facts that $\delta t_\gw$ will  typically be zero for a significant fraction of $T_\obs$, and that a large part of the effect will be absorbed in the pulsars' timing models (and especially by the fitting of their periods and period derivatives) \cite{2010MNRAS.401.2372V}.
Note that GW memory effect is essentially a low-frequency effect: SNR can build up precisely because memory remains constant, but non-zero, for a sizable fraction of $T_{obs}$. Thus there is no particular advantage to high-cadence timing measurements.

We can now derive how large an SNR we may expect for detecting GW memory for a given $\Omega_\gw$ and for given observational parameters. As above, we relate the energy density in GWs to the energy emitted in GW bursts, 
\begin{equation}
\label{eq:mem4}
\Omega_{\gw} \sim 10 \, \Delta E \, R_4\, \tau^3_0;
\end{equation}
we then combine Eqs.\ \eqref{eq:mem1}, \eqref{eq:mem-snr}, and set $d = d_\mathrm{near} = (4 \pi R_4\, T_{\obs} / 3)^{-1/3}$, to obtain
\begin{equation}
\label{eq:mem5}
\snr_\mathrm{mem,near} \sim \frac{\alpha}{300} \, \frac{\Omega_{\gw}}{\tau_0^3} R_4^{-2/3} T_{\obs}^{4/3} \frac{(M p \, T_{\obs})^{1/2}}{\delta t_{\rms}}.
\end{equation}
Again, for fixed $\Omega_\gw$ we maximize $\snr_\mathrm{mem,near}$ by taking $R_4$ to be as small as possible, subject to the constraint that $d_\mathrm{near} < \tau_0$, leading to 
\begin{equation}
\label{eq:snrmaxmem}
\begin{aligned}
\max\{\snr_\mathrm{mem} \}& \simeq \frac{\alpha}{500} \, \frac{\Omega_{\gw}}{\tau_0}
T^2_{\obs} \frac{(M p \, T_{\obs})^{1/2}}{\delta t_\rms} \\
& \simeq 700 \, \alpha \, \bigg[\frac{\Omega_{\gw}}{10^{-2}}\bigg] 
\bigg[\frac{T_{\obs}}{10^8 \, \mathrm{s}}\bigg]^2 \times \mathrm{obs.}
\end{aligned}
\end{equation}

Comparing Eqs.~(\ref{eq:fid1}) and (\ref{eq:snrmaxmem}), we see that--depending on the 
values of $\Omega_{\gw}$ and $f$--the memory effect from a burst could  be much more detectable than its direct waves.  
More generally, comparing $\snr_\mem$ with the \emph{direct} SNR for the same source, as given by Eqs.\ \eqref{eq:snr2} and \eqref{eq:dt}, we find:
\newcommand{\dir}{\mathrm{dir}}
\begin{equation}
\label{eq:ratio_lowz}
\begin{aligned}
\frac{\snr_\mem}{\snr_\dir} &= \frac{1/20}{1/20} \frac{h_\mem T_\obs}{h / f} \left(\frac{M p \, T_\obs}{M p \, T_\sig}\right)^{1/2} \\
&= \frac{1/20}{1/20} \frac{\pi^2 \alpha}{2 \sqrt{6}} h \, f^3 \, d \, T_\sig \, T_\obs \left(\frac{M p \, T_\obs}{M p \,T_\sig}\right)^{1/2} \\
&= \frac{1}{1/20} \frac{\pi^2 \alpha}{2 \sqrt{6}} \, \snr_\dir \, \frac{\delta t_\rms \, T_\sig^{-4} \, T_\obs^2 \, d}{(M p \,T_\obs)^{1/2}} \\
&\simeq 10^6 \, \alpha \, \snr_\dir
\bigg[\frac{T_\sig}{10^5 \, \mathrm{s}} \bigg]^{-4}
\bigg[\frac{T_\obs}{10^8 \, \mathrm{s}} \bigg]^{2}
\bigg[\frac{d}{\tau_0} \bigg]
\big[\mathrm{obs.} \big]^{-1}
\end{aligned}
\end{equation}
where in the second row we have used the fact that $h_\mem \simeq (\alpha / \sqrt{6}) (\Delta E/d)$ and $\Delta E = (\pi^2/2) h^2 f^2 d^2 \times T_\sig$; in the third row we have substituted $\snr_\dir = (1/20) (h/f) \delta t_\rms^{-1} (M p T_\sig)^{1/2}$ and replaced $f$ with $1/T_\sig$, as appropriate for a burst signal. 
Since $\snr_\mathrm{dir}$  scales as $h_{\dir}$ while $\snr_\mathrm{mem}$ scales as $h^2_{\dir}$, the memory effect dominates for a sufficiently strong signal.

\section{Discovery space at high redshift}
\label{sec:highredshift}
In the previous section we have considered sources at small $z$, neglecting cosmological effects. We now turn to sources in the early Universe, at $z \gg 1$.   
Again, we will assume that the sources are isotropically distributed and that the Earth does not have a preferred location in spacetime with respect to them. The especially interesting cases are GW memory, which we discuss first, and unmodeled bursts.

We begin by collecting a few useful formulas. Let $t \equiv \int a^{-1}(\tau) \,\mathrm{d}\tau$ be the conformal time coordinate, in terms of which the (spatially flat) Robertson--Walker metric becomes
\be
ds^2 = a^2(t)\big[-dt^2 + dx^2 + dy^2 + dz^2\big] \,.
\ee
\newcommand{\eq}{\mathrm{eq}}
We find it useful to define the high-$z$ epoch into the radiation-dominated era for $z \ll z_\eq$ and the matter-dominated era for $z \gg z_\eq$, where $z_\eq \approx 3,200$ (the redshift at which the energies of matter and radiation were equal). Then we can approximate $a(\tau) \propto \tau^{1/2}$ for $\tau < \tau_\eq$ and 
$a(\tau) \propto \tau^{2/3}$ for $\tau > \tau_\eq$ (of course, we now know that the Universe is dark-energy, rather than matter dominated for $z \alt 1.7$, but we neglect this correction in keeping with the back-of-the-envelope spirit of this paper).

We use the subscript ``0'' to refer to present Universe (e.g., $\tau_0 \sim 10^{10}$ years is the present age of the Universe), and we choose our spatial coordinates so that $a_0 \equiv a(\tau_0) = 1$.  Then 
\be\label{tz}
 t(z) = \begin{cases} (1 + z_{\eq})\big(3 \tau_{\eq}^{2/3}\,\tau^{1/3}(z) - \tau_{\eq}\big) & z < z_{\eq}, \\
(1 + z_{\eq})\big(2 \tau_{\eq}^{1/2}\,\tau^{1/2}(z) \big) & z > z_{\eq}, \, .
\end{cases}
\ee
and in particular, 
\be\label{t0}
t_0 \simeq (1 + z_{\eq})\big(3 \tau_{\eq}^{2/3}\,\tau_0^{1/3} \big) \,,
\ee
and therefore
\be\label{t0tz}
\frac{t_0}{t(z)} \simeq \begin{cases} (1 + z)^{1/2} & z < z_{\eq}, \\
\frac{3}{2} (1+z) (1 + z_{\eq})^{-1/2} & z > z_{\eq}, \, .
\end{cases}
\ee

Now consider GW bursts produced at $z \gg 1$.  The size of the particle horizon at redshift $z$ is $\sim t(z)$ in co-moving coordinates, and so the number of such particle-horizon volumes within our horizon volume today is $\sim [t_0/t(z)]^3$.  Let $B$ be the average number of GW bursts coming from each horizon volume $[t(z)]^3$.
Let the energy (as measured at $z$) of a typical burst $\Delta E(z)$; by today that energy has been redshifted to $\Delta E_0 = \Delta E_z/(1+z)$. The total energy today, within a Hubble volume, from all such bursts at redshift $z$ is $\Delta E_0 \,B\, [t_0/t(z)]^3$, and it satisfies
\be\label{DelE_Omeg}
\Delta E_0 \,B\, [t_0/t(z)]^3 \lesssim \frac{1}{10} \Omega_{\gw} \tau_0 \, .
\ee
We write ``$\lesssim$'' instead of ``$\simeq$'' because there could be other significant sources for $\Omega_{\gw}$, besides this early-Universe contribution.

\subsection{Discovery space for GW memory from sources at high $z$}
\label{sec:zlarge}
The generalization of Eq.\ \eqref{eq:mem1} to sources at arbitrary $z$ is
\begin{equation}
h_{\mem} \sim \frac{\alpha}{\sqrt{6}} \frac{\Delta E(z) (1+z)}{D_L},
\end{equation}
where $\Delta E(z)$ is the locally measured energy loss and $D_L$ is the luminosity distance to the source (this follows from the propagation of GW-like perturbations in the Robertson--Walker spacetime \cite{maggiore2008} and from the definition of $D_L$). The energy carried by those emitted waves today is $\Delta E_0 = \Delta E(z)/(1+z)$, while for high $z$ we have $D_L \approx 3\tau_0 (1+z)$.
Thus we have
\be\label{hmemz}
h_{\mem} \simeq \frac{\alpha}{8} \frac{\Delta E_0\,(1+z)}{\tau_0} \,.
\ee
It is instructive to determine the high-$z$ version of Eq.\ \eqref{eq:ratio_lowz} for the ratio $\snr_{\mem}/\snr_{\dir}$. The only change in the derivation is the replacement $d \rightarrow 3 \tau_0 (1+z)$, leading to: 
\begin{equation}
\begin{aligned}
\frac{\snr_{\mem}}{\snr_{\dir}} 
\simeq & \, 3 \times 10^{13} \, \alpha \, \snr_\dir \\
& \times \bigg[\frac{1+z}{10^7}\bigg]
\bigg[\frac{T_\sig}{10^5 \, \mathrm{s}} \bigg]^{-4}
\bigg[\frac{T_\obs}{10^8 \, \mathrm{s}} \bigg]^{2}
\bigg[\frac{d}{\tau_0} \bigg]
\big[\mathrm{obs.} \big]^{-1}
\end{aligned}
\end{equation}

By combining Eqs.~(\ref{t0tz}), \eqref{DelE_Omeg}, and \eqref{hmemz}, we can constrain $\snr_{\mem}$ given $B$ and $\Omega_\gw$:
\be
h_{\mem} \alt \frac{\alpha}{80} \frac{\Omega_{\gw}}{B} \times \begin{cases} \frac{1}{(1+z)^{1/2}} & 1 \ll z \ll z_{\eq}, \\
\frac{(1+ z_{\eq})^{3/2}}{3 (1+z)^{2}} & z \gg z_{\eq} \, ; \end{cases}
\ee
the corresponding SNR follows from Eq.\ \eqref{eq:mem-snr}. We want to have a high probability of seeing one such signal within the observation time $T_\obs$. Since the local rate can be shown~\footnote{Briefly, this can be shown by using Eq.~(10) of \cite{2006PhRvD..73d2001C}, approximating the term $4\pi (a_0 r_1)^2 \equiv  
4\pi \big(a_0 (t_0 - t(z) ) \big)^2$ by  
$4\pi (a_0 t_0)^2 \equiv 4\pi (\tau_0)^2$ and using $\dot n(z) (d\tau_1/dz)\Delta z =
\dot n(z) \Delta \tau_1 = (B/\tau^3_0)(t_0/t(z))^3$.}
to be $R \sim 4 \pi (B/\tau_0) [t_0/t(z)]^3$.  Imposing $R \, T_\obs \gtrsim 1$ leads to
\begin{equation}
\begin{aligned}
\max\{\snr_\mathrm{mem}\} &\simeq \frac{\alpha}{125} (1+z) \frac{\Omega_\gw}{\tau_0} T^2_{\obs} \frac{(M p \, T_{\obs})^{1/2}}{\delta t_{\rms}} \\
& \simeq 270 \, \alpha \, \bigg[\frac{1+z}{10^7}\bigg] \bigg[\frac{\Omega_{\gw}}{10^{-10}}\bigg] 
\bigg[\frac{T_{\obs}}{10^8 \, \mathrm{s}}\bigg]^2 \times \mathrm{obs.},
\end{aligned}
\end{equation}
a factor of order $(1+z)$ larger than the limit we derived in Eq.\ \eqref{eq:snrmaxmem} for sources at $z \alt 1$. We regard this as a promising result, since current constraints on $\Omega_{\gw}$ still leave a great deal of room for possible discovery.

\subsection{Discovery space for unmodeled GW bursts at high $z$}

We now examine the prospects for detecting a GW burst from high $z$. The total energy emitted by such a source is 
\be \label{eq:dirz}
\Delta E(z) = \Delta E_0 \,(1+z) \simeq
\frac{\pi^2}{2} h^2 f^2 T_{\sig} D^2_L;
\ee
using Eq.\ \eqref{DelE_Omeg} and $D_L \sim 3 \tau_0 (1+z)$, we then have
\begin{equation}
\begin{aligned} 
h^2 \lesssim & \,\, 2\times 10^{-3}\ \frac{\Omega_{\gw}}{B} (f \, \tau_0)^{-1} (f \, T_{\sig})^{-1} \\
& \times \begin{cases} (1+z)^{-5/2} & 1 \ll z \ll z_{\eq}, \\
(1/3)  (1+ z_{\eq})^{3/2}\, (1+z)^{-4} & z \gg z_{\eq} \, . \end{cases}
\end{aligned}
\end{equation}
Again, a high probability of observing a signal constrains the rate $R$ according to
$R \max\{T_\sig,T_\obs\} \agt 1$, leading to 
\begin{equation}
\label{eq:dirhighz}
\begin{aligned}
\max\{\snr_\mathrm{dir} \} & \lesssim \frac{1}{120}
\bigg[\frac{\Omega_{\gw}}{1+z}\bigg]^{1/2} 
\frac{\big(M p \, T_{\obs}\big)^{1/2}}{(f\,\delta t_{\rms})(f\, \tau_0)} \\
& \approx 10
\bigg[\frac{f}{10^{-7} \, \mathrm{Hz}}\bigg]^{-2}
\bigg[\frac{\Omega_\gw}{10^{-5} (1+z)}\bigg]^{1/2}
\times \mathrm{obs.}
\end{aligned}
\end{equation}
This is basically the same limit we found for the largest-SNR burst at $z < 1$, but
multiplied by the factor $(1+z)^{-1/2}$.

\section{Corrections for beaming and for Galactic sources}
\label{sec:galactic}
So far our estimates of signal strengths have implicitly assumed that the radiation is not strongly beamed.  We have also implicitly assumed that detectable PTA sources will be extra-Galactic. In this section we briefly show how our estimates get modified if one drops these assumptions.  Both these issues were addressed by ZT82 \cite{Zimmermann:1982wi}, but here we extend their considerations to large $z$.

\subsection{Modifications for highly beamed radiation}  
Assume that the GW energy is beamed into solid angle $4\pi F$.  To see how $\max\{\snr\}$ for ``direct'' radiation scales with $F$, we will take $\Omega_{\gw}$ and the total radiated energy to be fixed, which together imply a fixed rate density.  For the case $z \alt 1$,  we can approximate space as Euclidean, so the distance $d$ to the closest source beaming in our direction scales as $d \propto F^{-1/3}$; the observed $h$ scales as $h \propto F^{-1/2}/d$; and altogether $h \propto F^{-1/6}$.
We see that the effect of beaming on $\max\{\snr\}$ is extremely weak; for instance, a beaming factor $F = 10^{-3}$ yields only a factor $\sim 3$ increase in the potential SNR. This very weak dependence was already noted by ZT82 in the $z \alt 3$ case.

For $z \gg 1$, to account for beaming, on the right hand side of Eq.~\eqref{eq:dirz} we 
would replace $\Delta E_0$ with $\Delta E_0/F$.  However the condition $R \, T_{\obs} \agt 1 $ gets replaced by  $R \, F \, T_{\obs} \agt 1 $, which leads to $\Delta E_0 \propto \Omega_{\gw} B^{-1} F$.  Thus the $F$ factors cancel, and beaming has basically no effect on $\max\{\snr\}$ for high-$z$ sources.
Note that our low-$z$ and high-$z$ upper limits, Eqs.\ \eqref{eq:fid1} and \eqref{eq:dirhighz} respectively, have slightly different character: for the former we maximize the SNR from the nearest detected source, for the latter we fix $z$ and therefore luminosity distance under the constraint of detecting at least one source during the experiment.

What about memory? The effect of beaming is negligible, since the memory component of GW strain is not beamed, even when direct waves are. The dominant effect is that the parameter $\alpha$ changes by a factor of order one compared to the case of quadrupole emission. \red{MV: need ref?}

\subsection{Modifications for Galactic sources}
\label{galactic} 

Throughout Secs.\ \ref{sec:z1} and \ref{sec:highredshift} we have assumed that the Earth does not occupy a preferred location in the Universe.  
However the Earth lies in the Galaxy; how might that modify our results?
For sources at low $z$,  universe,  we showed in Sec.\ \ref{sec:gwsig} that, for fixed $\Omega_{\gw}$, detection SNR is maximized for sources whose event rate is once per $T_{\obs}$ in a Hubble volume.  For a Galactic source to be observable, this rate must increase to once per $T_{\obs}$ per Milky-Way-like galaxy, or $\sim 10^9$ times greater. To maintain the same $\Omega_{\gw}$, the energy $\Delta E$ radiated per event must decrease by a factor $10^9$. (We must also assume that the Galaxy can sustain such a rate of events.) On the other hand, the distance to the extra-Galactic source is $\sim 3$ Gpc, compared to $\sim 10$ kpc for a randomly located Galactic source.  For the direct radiation, $h \propto \Delta E^{1/2}/d$, so the ratio
\begin{equation}
\frac{\max\{\snr_{\mathrm{dir}}^{\mathrm{Gal}}\}}{\max\{\snr_{\mathrm{dir}}^{z \sim 1}\}} \sim 10^{-9/2} \frac{3 \, \mathrm{Gpc}}{10 \, \mathrm{kpc}} \sim 10,
\end{equation}
as was first shown by ZT82 \cite{Zimmermann:1982wi}.   Thus, besides being intrinsically less plausible, putative Galactic sources increase $\max\{\snr_{\mathrm{dir}}\}$ by only an order of magnitude, compared to the $z \sim 1$ case.

While we have undertaken the above calculation in the spirit of completeness, we point out that to account for an overall $\Omega_{\gw} \sim 10^{-2}$ (say), these putative Galactic explosions would have to release $\sim 50 M_{\odot}$ in GW energy roughly every $\sim 3\,$yr, and it would appear difficult to construct a plausible physical mechanism for such explosions that would not already have been detected by other means.

For the memory effect, $h \propto \Delta E/d$, so we may estimate a ratio
\begin{equation}
\frac{\max\{\snr_{\mathrm{mem}}^{\mathrm{Gal}}\}}{\max\{\snr_{\mathrm{mem}}^{z \sim 1}\}} \sim 10^{-9/2} \frac{3 \, \mathrm{Gpc}}{10 \, \mathrm{kpc}} \sim 10^{-3.5}.
\end{equation}

Finally, we note that if we had focused on sources in the Local Group instead of just the Milky Way, the event rate for sources outside the Milky Way would be dominated by Andromeda.  Since Andromeda has roughly the same mass as the Milky Way but is $\sim 100$ times further away than our Galactic Center, the strongest such events would be
$\sim 100$ times weaker than Galactic events.

\section{Conclusions and  Caveats}
\label{sec:summ}
In the paper we have constrained and characterized the GW discovery space of PTAs on the basis of energetic and statistical considerations alone.
In Secs.\ \ref{sec:z1} and \ref{sec:highredshift} we showed that a PTA detection of GWs at frequencies above $\sim 3 \times 10^{-5}$ Hz would either be an extraordinary coincidence, or have extraordinary implications; this effect results from an analysis of 
fundamental constraints on possible sources across the PTA sensitivity range, rather than deficiencies in PTA detection itself.
We showed also that GW memory can be more detectable than direct GWs, and that memory increasingly dominates the total SNR of an event for sources at higher and higher redshifts; indeed, GW memory from high-$z$ sources represents a large discovery space for PTAs.

Although we assumed modest beaming in our estimates, in Sec.\ \ref{sec:galactic} we argued that even extreme beaming would have a minor impact on detection SNRs. 
Similarly, although we assumed that the strongest GW sources during PTA observation would be extragalactic, our constraint on $\max\{\snr\}$ rises only by a factor $\sim 10$ for Galactic sources. Throughout the paper we adopted an SNR scaling law valid for white pulsar noise; in Sec.\ \ref{sec:determ} we explained, on the basis of toy model and of the observational characterization of pulsar noise, why this was appropriate.

In Sec.\ \ref{sec:degen} we demonstrated how to properly incorporate the effects of red noise in PTA searches, and we demonstrated that the effects of periodic GWs between $\sim 10^{-7.5}$ and $10^{-4.5}\,$ Hz band would \emph{not} be degenerate with small errors in the standard pulsar parameters, except in a few very narrow bands.

Theoretical upper limits are akin to no-go theorems, and the authors are well aware that the history of the latter in physics is replete with examples of results that, while strictly correct, turned out to be misleading because their assumptions were overly restrictive.
%assumptions were overly restrictive.~\footnote{E.g., the %Coleman-Mandula theorem famously restricted the possible %symmetries of relativistic particle physics--but did not account for %the possibility of symmetries that mixed bosons with fermions, %and so missed the possibility of supersymmetry. Similarly, the .... %classification of possible crystal symmetries did not account for %the possibility of quasi-crystals.}  
For this reason, our chief motivation in doing this research was not to rule out possibilities, but to uncover promising but neglected areas of search space.  
With this in mind, we now recall some of the assumptions that we have made, and point out some of the ways that Nature could be side-stepping them.
\begin{itemize}
\item In this paper we assumed that the Earth is not in a preferred location in the Universe. In Sec.~\ref{sec:galactic} we considered the case in which relevant GW sources are clustered in galaxies, but still assumed that the Earth is not in some preferred location within the Milky Way.
\item Even if the Earth does not occupy a preferred location with respect to relevant GW sources, some millisecond pulsars might do so. For instance, if two or more pulsars are located in a globular cluster that \emph{also} contains a BH binary with masses $\agt 1000 M_{\odot}$ 
%and period less than a few days \red{[check]}, 
% SBS: if the binary is massive enough it does not matter what the period is.
the correlated timing residuals due to the binary's GWs impinging on the pulsars could well be detectable (see, e.g., \cite{2005ApJ...627L.125J}).
%CC: The numerical result for a pair 1000Msun BHs with f (=2 \nu_orb) = 10^{-7} Hz
%is h = 1.3 

%
\item In this paper we assumed that at any redshift $z$ there are no structures (such as phase-transition bubbles) that are significantly larger than the contemporaneous horizon size $t(z)$. This is a reasonable way to incorporate causality constraints for processes that are not correlated on super-horizon scales to begin with, but it certainly does not hold for all cases: for instance, inflation would imprint correlations on much larger scales. So \emph{a priori} there could arise strong GW sources that violate this assumption.
\end{itemize}
It might be worthwhile to try to come up with reasonable physical scenarios that violate one or more of our assumptions.

\begin{acknowledgments}
CC gratefully acknowledges support from NSF Grant PHY-1068881. MV is grateful for support from the JPL RTD program.
This work was carried out at the Jet Propulsion Laboratory, California Institute of Technology, under contract to the National Aeronautics and Space Administration. Copyright 2013 California Institute of Technology.
\end{acknowledgments}

%\appendix
%\section{Derivation of Eq.~(\ref{dnear}) for $d_{near}$.}
%Here we briefly justify Eq.~(\ref{dnear}), which is equivalent to
%\be\label{dnear_app}
%d_{near} \sim \begin{cases} \bigg[\frac{3}{4\pi} V_R \, \frac{T_R}{T_{obs}} \bigg]^{1/3}\, &T_{sig} << T_{obs} \\
%\bigg[ \frac{3}{4\pi} V_R \, \frac{T_R}{T_{sig}} \bigg]^{1/3} &T_{sig} << T_{obs} \, . \end{cases}
%\ee
%
%This is not an exact formula; rather it is a reasonable approximation that interpolates smoothly between the limits $T_{sig} << T_{obs}$ and  $T_{sig} >> T_{obs}$. 
%First consider the case $T_{sig} << T_{obs}$  Define $V_{near} \equiv 1/(R_4 T_{obs})$ (where we recall $1/R_4 \equiv V_R  T_R$).
%Then probability of having zero events in some volume $V$ during $T_{obs}$ is clearly $exp(-V/V_{near}$, so $1-e^{-1} \approx  0.63$ of the time the closest event is within $d_{near} = \frac{3}{4\pi} V_{near}\big]^{1/3}$.
%At the other extreme, $T_{sig} >> T_{obs}$, the spatial density of observed events clearly increases by a factor  $T_{sig}/T_{obs}$, and so  $d_{near}$ is reduced by a factor $(T_{sig}/T_{obs})^{1/3}$.  This is precisely what is reflected in Eq.~(\ref{dnear_app}).

\bibliography{PTA-estimates}

\def\eprinttmppp@#1arXiv:@{#1}
\providecommand{\arxivlink[1]}{\href{http://arxiv.org/abs/#1}{arXiv:#1}}
\def\eprinttmp@#1arXiv:#2 [#3]#4@{\ifthenelse{\equal{#3}{x}}{\ifthenelse{
\equal{#1}{}}{\arxivlink{\eprinttmppp@#2@}}{\arxivlink{#1}}}{\arxivlink{#2}
  [#3]}}
\providecommand{\eprintlink}[1]{\eprinttmp@#1arXiv: [x]@}
\renewcommand{\eprint}[1]{\eprintlink{#1}}
\providecommand{\adsurl}[1]{\href{#1}{ADS}}
\renewcommand{\bibinfo}[2]{\ifthenelse{\equal{#1}{isbn}}{\href{http://cosmologist.info/ISBN/#2}{#2}}{#2}}
\begin{thebibliography}{46}
\expandafter\ifx\csname natexlab\endcsname\relax\def\natexlab#1{#1}\fi
\expandafter\ifx\csname bibnamefont\endcsname\relax
  \def\bibnamefont#1{#1}\fi
\expandafter\ifx\csname bibfnamefont\endcsname\relax
  \def\bibfnamefont#1{#1}\fi
\expandafter\ifx\csname citenamefont\endcsname\relax
  \def\citenamefont#1{#1}\fi
\expandafter\ifx\csname url\endcsname\relax
  \def\url#1{\texttt{#1}}\fi
\expandafter\ifx\csname urlprefix\endcsname\relax\def\urlprefix{URL }\fi

\bibitem[{\citenamefont{{Sazhin}}(1978)}]{1978SvA....22...36S}
\bibinfo{author}{\bibfnamefont{M.~V.} \bibnamefont{{Sazhin}}},
  \bibinfo{journal}{\sovast} \textbf{\bibinfo{volume}{22}}, \bibinfo{pages}{36}
  (\bibinfo{year}{1978}),
  \adsurl{http://adsabs.harvard.edu/abs/1978SvA....22...36S}.

\bibitem[{\citenamefont{{Detweiler}}(1979)}]{1979ApJ...234.1100D}
\bibinfo{author}{\bibfnamefont{S.}~\bibnamefont{{Detweiler}}},
  \bibinfo{journal}{\apj} \textbf{\bibinfo{volume}{234}}, \bibinfo{pages}{1100}
  (\bibinfo{year}{1979}),
  \adsurl{http://adsabs.harvard.edu/abs/1979ApJ...234.1100D}.

\bibitem[{\citenamefont{{Hellings} and {Downs}}(1983)}]{1983ApJ...265L..39H}
\bibinfo{author}{\bibfnamefont{R.~W.} \bibnamefont{{Hellings}}}
  \bibnamefont{and} \bibinfo{author}{\bibfnamefont{G.~S.}
  \bibnamefont{{Downs}}}, \bibinfo{journal}{\apjl}
  \textbf{\bibinfo{volume}{265}}, \bibinfo{pages}{L39} (\bibinfo{year}{1983}),
  \adsurl{http://adsabs.harvard.edu/abs/1983ApJ...265L..39H}.

\bibitem[{\citenamefont{{van Haasteren} et~al.}(2011)\citenamefont{{van
  Haasteren}, {Levin}, {Janssen}, {Lazaridis}, {Kramer}, {Stappers},
  {Desvignes}, {Purver}, {Lyne}, {Ferdman} et~al.}}]{2011MNRAS.414.3117V}
\bibinfo{author}{\bibfnamefont{R.}~\bibnamefont{{van Haasteren}}},
  \bibinfo{author}{\bibfnamefont{Y.}~\bibnamefont{{Levin}}},
  \bibinfo{author}{\bibfnamefont{G.~H.} \bibnamefont{{Janssen}}},
  \bibinfo{author}{\bibfnamefont{K.}~\bibnamefont{{Lazaridis}}},
  \bibinfo{author}{\bibfnamefont{M.}~\bibnamefont{{Kramer}}},
  \bibinfo{author}{\bibfnamefont{B.~W.} \bibnamefont{{Stappers}}},
  \bibinfo{author}{\bibfnamefont{G.}~\bibnamefont{{Desvignes}}},
  \bibinfo{author}{\bibfnamefont{M.~B.} \bibnamefont{{Purver}}},
  \bibinfo{author}{\bibfnamefont{A.~G.} \bibnamefont{{Lyne}}},
  \bibinfo{author}{\bibfnamefont{R.~D.} \bibnamefont{{Ferdman}}},
  \bibnamefont{et~al.}, \bibinfo{journal}{\mnras}
  \textbf{\bibinfo{volume}{414}}, \bibinfo{pages}{3117} (\bibinfo{year}{2011}),
  \eprint{1103.0576}.

\bibitem[{\citenamefont{{Demorest} et~al.}(2013)\citenamefont{{Demorest},
  {Ferdman}, {Gonzalez}, {Nice}, {Ransom}, {Stairs}, {Arzoumanian}, {Brazier},
  {Burke-Spolaor}, {Chamberlin} et~al.}}]{2013ApJ...762...94D}
\bibinfo{author}{\bibfnamefont{P.~B.} \bibnamefont{{Demorest}}},
  \bibinfo{author}{\bibfnamefont{R.~D.} \bibnamefont{{Ferdman}}},
  \bibinfo{author}{\bibfnamefont{M.~E.} \bibnamefont{{Gonzalez}}},
  \bibinfo{author}{\bibfnamefont{D.}~\bibnamefont{{Nice}}},
  \bibinfo{author}{\bibfnamefont{S.}~\bibnamefont{{Ransom}}},
  \bibinfo{author}{\bibfnamefont{I.~H.} \bibnamefont{{Stairs}}},
  \bibinfo{author}{\bibfnamefont{Z.}~\bibnamefont{{Arzoumanian}}},
  \bibinfo{author}{\bibfnamefont{A.}~\bibnamefont{{Brazier}}},
  \bibinfo{author}{\bibfnamefont{S.}~\bibnamefont{{Burke-Spolaor}}},
  \bibinfo{author}{\bibfnamefont{S.~J.} \bibnamefont{{Chamberlin}}},
  \bibnamefont{et~al.}, \bibinfo{journal}{\apj} \textbf{\bibinfo{volume}{762}},
  \bibinfo{eid}{94} (\bibinfo{year}{2013}), \eprint{1201.6641}.

\bibitem[{\citenamefont{{Manchester} et~al.}(2013)\citenamefont{{Manchester},
  {Hobbs}, {Bailes}, {Coles}, {van Straten}, {Keith}, {Shannon}, {Bhat},
  {Brown}, {Burke-Spolaor} et~al.}}]{2013PASA...30...17M}
\bibinfo{author}{\bibfnamefont{R.~N.} \bibnamefont{{Manchester}}},
  \bibinfo{author}{\bibfnamefont{G.}~\bibnamefont{{Hobbs}}},
  \bibinfo{author}{\bibfnamefont{M.}~\bibnamefont{{Bailes}}},
  \bibinfo{author}{\bibfnamefont{W.~A.} \bibnamefont{{Coles}}},
  \bibinfo{author}{\bibfnamefont{W.}~\bibnamefont{{van Straten}}},
  \bibinfo{author}{\bibfnamefont{M.~J.} \bibnamefont{{Keith}}},
  \bibinfo{author}{\bibfnamefont{R.~M.} \bibnamefont{{Shannon}}},
  \bibinfo{author}{\bibfnamefont{N.~D.~R.} \bibnamefont{{Bhat}}},
  \bibinfo{author}{\bibfnamefont{A.}~\bibnamefont{{Brown}}},
  \bibinfo{author}{\bibfnamefont{S.~G.} \bibnamefont{{Burke-Spolaor}}},
  \bibnamefont{et~al.}, \bibinfo{journal}{\pasa} \textbf{\bibinfo{volume}{30}},
  \bibinfo{eid}{e017} (\bibinfo{year}{2013}), \eprint{1210.6130}.

\bibitem[{\citenamefont{{Hobbs} et~al.}(2010)\citenamefont{{Hobbs},
  {Archibald}, {Arzoumanian}, {Backer}, {Bailes}, {Bhat}, {Burgay},
  {Burke-Spolaor}, {Champion}, {Cognard} et~al.}}]{2010CQGra..27h4013H}
\bibinfo{author}{\bibfnamefont{G.}~\bibnamefont{{Hobbs}}},
  \bibinfo{author}{\bibfnamefont{A.}~\bibnamefont{{Archibald}}},
  \bibinfo{author}{\bibfnamefont{Z.}~\bibnamefont{{Arzoumanian}}},
  \bibinfo{author}{\bibfnamefont{D.}~\bibnamefont{{Backer}}},
  \bibinfo{author}{\bibfnamefont{M.}~\bibnamefont{{Bailes}}},
  \bibinfo{author}{\bibfnamefont{N.~D.~R.} \bibnamefont{{Bhat}}},
  \bibinfo{author}{\bibfnamefont{M.}~\bibnamefont{{Burgay}}},
  \bibinfo{author}{\bibfnamefont{S.}~\bibnamefont{{Burke-Spolaor}}},
  \bibinfo{author}{\bibfnamefont{D.}~\bibnamefont{{Champion}}},
  \bibinfo{author}{\bibfnamefont{I.}~\bibnamefont{{Cognard}}},
  \bibnamefont{et~al.}, \bibinfo{journal}{Classical and Quantum Gravity}
  \textbf{\bibinfo{volume}{27}}, \bibinfo{eid}{084013} (\bibinfo{year}{2010}),
  \eprint{0911.5206}.

\bibitem[{\citenamefont{{Sesana}}(2013{\natexlab{a}})}]{2013MNRAS.433L...1S}
\bibinfo{author}{\bibfnamefont{A.}~\bibnamefont{{Sesana}}},
  \bibinfo{journal}{\mnras} \textbf{\bibinfo{volume}{433}}, \bibinfo{pages}{L1}
  (\bibinfo{year}{2013}{\natexlab{a}}), \eprint{1211.5375}.

\bibitem[{\citenamefont{{Sesana} et~al.}(2009)\citenamefont{{Sesana},
  {Vecchio}, and {Volonteri}}}]{2009MNRAS.394.2255S}
\bibinfo{author}{\bibfnamefont{A.}~\bibnamefont{{Sesana}}},
  \bibinfo{author}{\bibfnamefont{A.}~\bibnamefont{{Vecchio}}},
  \bibnamefont{and}
  \bibinfo{author}{\bibfnamefont{M.}~\bibnamefont{{Volonteri}}},
  \bibinfo{journal}{\mnras} \textbf{\bibinfo{volume}{394}},
  \bibinfo{pages}{2255} (\bibinfo{year}{2009}), \eprint{0809.3412}.

\bibitem[{\citenamefont{{Yardley} et~al.}(2010)\citenamefont{{Yardley},
  {Hobbs}, {Jenet}, {Verbiest}, {Wen}, {Manchester}, {Coles}, {van Straten},
  {Bailes}, {Bhat} et~al.}}]{2010MNRAS.407..669Y}
\bibinfo{author}{\bibfnamefont{D.~R.~B.} \bibnamefont{{Yardley}}},
  \bibinfo{author}{\bibfnamefont{G.~B.} \bibnamefont{{Hobbs}}},
  \bibinfo{author}{\bibfnamefont{F.~A.} \bibnamefont{{Jenet}}},
  \bibinfo{author}{\bibfnamefont{J.~P.~W.} \bibnamefont{{Verbiest}}},
  \bibinfo{author}{\bibfnamefont{Z.~L.} \bibnamefont{{Wen}}},
  \bibinfo{author}{\bibfnamefont{R.~N.} \bibnamefont{{Manchester}}},
  \bibinfo{author}{\bibfnamefont{W.~A.} \bibnamefont{{Coles}}},
  \bibinfo{author}{\bibfnamefont{W.}~\bibnamefont{{van Straten}}},
  \bibinfo{author}{\bibfnamefont{M.}~\bibnamefont{{Bailes}}},
  \bibinfo{author}{\bibfnamefont{N.~D.~R.} \bibnamefont{{Bhat}}},
  \bibnamefont{et~al.}, \bibinfo{journal}{\mnras}
  \textbf{\bibinfo{volume}{407}}, \bibinfo{pages}{669} (\bibinfo{year}{2010}),
  \eprint{1005.1667}.

\bibitem[{\citenamefont{{Sesana}}(2013{\natexlab{b}})}]{2013arXiv1307.4086S}
\bibinfo{author}{\bibfnamefont{A.}~\bibnamefont{{Sesana}}},
  \bibinfo{journal}{ArXiv e-prints}  (\bibinfo{year}{2013}{\natexlab{b}}),
  \eprint{1307.4086}.

\bibitem[{\citenamefont{{van Haasteren} and
  {Levin}}(2010)}]{2010MNRAS.401.2372V}
\bibinfo{author}{\bibfnamefont{R.}~\bibnamefont{{van Haasteren}}}
  \bibnamefont{and} \bibinfo{author}{\bibfnamefont{Y.}~\bibnamefont{{Levin}}},
  \bibinfo{journal}{\mnras} \textbf{\bibinfo{volume}{401}},
  \bibinfo{pages}{2372} (\bibinfo{year}{2010}), \eprint{0909.0954}.

\bibitem[{\citenamefont{{Cordes} and {Jenet}}(2012)}]{2012ApJ...752...54C}
\bibinfo{author}{\bibfnamefont{J.~M.} \bibnamefont{{Cordes}}} \bibnamefont{and}
  \bibinfo{author}{\bibfnamefont{F.~A.} \bibnamefont{{Jenet}}},
  \bibinfo{journal}{\apj} \textbf{\bibinfo{volume}{752}}, \bibinfo{eid}{54}
  (\bibinfo{year}{2012}),
  \adsurl{http://adsabs.harvard.edu/abs/2012ApJ...752...54C}.

\bibitem[{\citenamefont{{Jaffe} and {Backer}}(2003)}]{2003ApJ...583..616J}
\bibinfo{author}{\bibfnamefont{A.~H.} \bibnamefont{{Jaffe}}} \bibnamefont{and}
  \bibinfo{author}{\bibfnamefont{D.~C.} \bibnamefont{{Backer}}},
  \bibinfo{journal}{\apj} \textbf{\bibinfo{volume}{583}}, \bibinfo{pages}{616}
  (\bibinfo{year}{2003}), \eprint{arXiv:astro-ph/0210148}.

\bibitem[{\citenamefont{{Jenet}
  et~al.}(2005{\natexlab{a}})\citenamefont{{Jenet}, {Hobbs}, {Lee}, and
  {Manchester}}}]{2005ApJ...625L.123J}
\bibinfo{author}{\bibfnamefont{F.~A.} \bibnamefont{{Jenet}}},
  \bibinfo{author}{\bibfnamefont{G.~B.} \bibnamefont{{Hobbs}}},
  \bibinfo{author}{\bibfnamefont{K.~J.} \bibnamefont{{Lee}}}, \bibnamefont{and}
  \bibinfo{author}{\bibfnamefont{R.~N.} \bibnamefont{{Manchester}}},
  \bibinfo{journal}{\apjl} \textbf{\bibinfo{volume}{625}},
  \bibinfo{pages}{L123} (\bibinfo{year}{2005}{\natexlab{a}}),
  \eprint{arXiv:astro-ph/0504458}.

\bibitem[{\citenamefont{Zimmermann and Thorne}(1980)}]{Zimmermann:1982wi}
\bibinfo{author}{\bibfnamefont{M.}~\bibnamefont{Zimmermann}} \bibnamefont{and}
  \bibinfo{author}{\bibfnamefont{K.}~\bibnamefont{Thorne}}, in
  \emph{\bibinfo{booktitle}{{Essays in General Relativity, a Festschrift for
  Abraham Taub}}}, edited by \bibinfo{editor}{\bibfnamefont{F.~J.}
  \bibnamefont{Tipler}} (\bibinfo{publisher}{Academic Press},
  \bibinfo{address}{New York}, \bibinfo{year}{1980}), pp.
  \bibinfo{pages}{139--155}.

\bibitem[{\citenamefont{{Lorimer}}(2008)}]{2008LRR....11....8L}
\bibinfo{author}{\bibfnamefont{D.~R.} \bibnamefont{{Lorimer}}},
  \bibinfo{journal}{Living Reviews in Relativity}
  \textbf{\bibinfo{volume}{11}}, \bibinfo{pages}{8} (\bibinfo{year}{2008}),
  \eprint{0811.0762}.

\bibitem[{\citenamefont{{Stairs}}(2003)}]{2003LRR.....6....5S}
\bibinfo{author}{\bibfnamefont{I.~H.} \bibnamefont{{Stairs}}},
  \bibinfo{journal}{Living Reviews in Relativity} \textbf{\bibinfo{volume}{6}},
  \bibinfo{pages}{5} (\bibinfo{year}{2003}), \eprint{arXiv:astro-ph/0307536}.

\bibitem[{\citenamefont{Maggiore}(2008)}]{maggiore2008}
\bibinfo{author}{\bibfnamefont{M.}~\bibnamefont{Maggiore}},
  \emph{\bibinfo{title}{Gravitational Waves: Volume 1: Theory and
  Experiments}}, Gravitational Waves (\bibinfo{publisher}{Oxford University
  Press}, \bibinfo{year}{2008}).

\bibitem[{\citenamefont{{Bretthorst}}(2001)}]{2001AIPC..567....1B}
\bibinfo{author}{\bibfnamefont{G.~L.} \bibnamefont{{Bretthorst}}}, in
  \emph{\bibinfo{booktitle}{American Institute of Physics Conference Series}}
  (\bibinfo{year}{2001}), vol. \bibinfo{volume}{567} of
  \emph{\bibinfo{series}{American Institute of Physics Conference Series}}, pp.
  \bibinfo{pages}{1--28}.

\bibitem[{\citenamefont{{van Haasteren} et~al.}(2009)\citenamefont{{van
  Haasteren}, {Levin}, {McDonald}, and {Lu}}}]{2009MNRAS.395.1005V}
\bibinfo{author}{\bibfnamefont{R.}~\bibnamefont{{van Haasteren}}},
  \bibinfo{author}{\bibfnamefont{Y.}~\bibnamefont{{Levin}}},
  \bibinfo{author}{\bibfnamefont{P.}~\bibnamefont{{McDonald}}},
  \bibnamefont{and} \bibinfo{author}{\bibfnamefont{T.}~\bibnamefont{{Lu}}},
  \bibinfo{journal}{\mnras} \textbf{\bibinfo{volume}{395}},
  \bibinfo{pages}{1005} (\bibinfo{year}{2009}), \eprint{0809.0791}.

\bibitem[{\citenamefont{{Cutler} and {Harms}}(2006)}]{2006PhRvD..73d2001C}
\bibinfo{author}{\bibfnamefont{C.}~\bibnamefont{{Cutler}}} \bibnamefont{and}
  \bibinfo{author}{\bibfnamefont{J.}~\bibnamefont{{Harms}}},
  \bibinfo{journal}{\prd} \textbf{\bibinfo{volume}{73}}, \bibinfo{eid}{042001}
  (\bibinfo{year}{2006}), \eprint{arXiv:gr-qc/0511092}.

\bibitem[{\citenamefont{Hager}(1989)}]{hager1989}
\bibinfo{author}{\bibfnamefont{W.~W.} \bibnamefont{Hager}},
  \bibinfo{journal}{SIAM Review} \textbf{\bibinfo{volume}{31}},
  \bibinfo{pages}{pp. 221} (\bibinfo{year}{1989}), ISSN
  \bibinfo{issn}{00361445},
  \urlprefix\url{http://www.jstor.org/stable/2030425}.

\bibitem[{\citenamefont{{Cordes} and {Shannon}}(2010)}]{2010arXiv1010.3785C}
\bibinfo{author}{\bibfnamefont{J.~M.} \bibnamefont{{Cordes}}} \bibnamefont{and}
  \bibinfo{author}{\bibfnamefont{R.~M.} \bibnamefont{{Shannon}}},
  \bibinfo{journal}{ArXiv e-prints}  (\bibinfo{year}{2010}),
  \eprint{1010.3785}.

\bibitem[{\citenamefont{{Shannon} and {Cordes}}(2010)}]{2010ApJ...725.1607S}
\bibinfo{author}{\bibfnamefont{R.~M.} \bibnamefont{{Shannon}}}
  \bibnamefont{and} \bibinfo{author}{\bibfnamefont{J.~M.}
  \bibnamefont{{Cordes}}}, \bibinfo{journal}{\apj}
  \textbf{\bibinfo{volume}{725}}, \bibinfo{pages}{1607} (\bibinfo{year}{2010}),
  \eprint{1010.4794}.

\bibitem[{\citenamefont{{Verbiest} et~al.}(2009)\citenamefont{{Verbiest},
  {Bailes}, {Coles}, {Hobbs}, {van Straten}, {Champion}, {Jenet}, {Manchester},
  {Bhat}, {Sarkissian} et~al.}}]{2009MNRAS.400..951V}
\bibinfo{author}{\bibfnamefont{J.~P.~W.} \bibnamefont{{Verbiest}}},
  \bibinfo{author}{\bibfnamefont{M.}~\bibnamefont{{Bailes}}},
  \bibinfo{author}{\bibfnamefont{W.~A.} \bibnamefont{{Coles}}},
  \bibinfo{author}{\bibfnamefont{G.~B.} \bibnamefont{{Hobbs}}},
  \bibinfo{author}{\bibfnamefont{W.}~\bibnamefont{{van Straten}}},
  \bibinfo{author}{\bibfnamefont{D.~J.} \bibnamefont{{Champion}}},
  \bibinfo{author}{\bibfnamefont{F.~A.} \bibnamefont{{Jenet}}},
  \bibinfo{author}{\bibfnamefont{R.~N.} \bibnamefont{{Manchester}}},
  \bibinfo{author}{\bibfnamefont{N.~D.~R.} \bibnamefont{{Bhat}}},
  \bibinfo{author}{\bibfnamefont{J.~M.} \bibnamefont{{Sarkissian}}},
  \bibnamefont{et~al.}, \bibinfo{journal}{\mnras}
  \textbf{\bibinfo{volume}{400}}, \bibinfo{pages}{951} (\bibinfo{year}{2009}),
  \eprint{0908.0244}.

\bibitem[{\citenamefont{{Begelman} et~al.}(1980)\citenamefont{{Begelman},
  {Blandford}, and {Rees}}}]{1980Natur.287..307B}
\bibinfo{author}{\bibfnamefont{M.~C.} \bibnamefont{{Begelman}}},
  \bibinfo{author}{\bibfnamefont{R.~D.} \bibnamefont{{Blandford}}},
  \bibnamefont{and} \bibinfo{author}{\bibfnamefont{M.~J.}
  \bibnamefont{{Rees}}}, \bibinfo{journal}{\nat}
  \textbf{\bibinfo{volume}{287}}, \bibinfo{pages}{307} (\bibinfo{year}{1980}),
  \adsurl{http://adsabs.harvard.edu/abs/1980Natur.287..307B}.

\bibitem[{\citenamefont{{Wyithe} and {Loeb}}(2003)}]{2003ApJ...590..691W}
\bibinfo{author}{\bibfnamefont{J.~S.~B.} \bibnamefont{{Wyithe}}}
  \bibnamefont{and} \bibinfo{author}{\bibfnamefont{A.}~\bibnamefont{{Loeb}}},
  \bibinfo{journal}{\apj} \textbf{\bibinfo{volume}{590}}, \bibinfo{pages}{691}
  (\bibinfo{year}{2003}), \eprint{arXiv:astro-ph/0211556}.

\bibitem[{\citenamefont{{McWilliams} et~al.}(2012)\citenamefont{{McWilliams},
  {Ostriker}, and {Pretorius}}}]{2012arXiv1211.4590M}
\bibinfo{author}{\bibfnamefont{S.~T.} \bibnamefont{{McWilliams}}},
  \bibinfo{author}{\bibfnamefont{J.~P.} \bibnamefont{{Ostriker}}},
  \bibnamefont{and}
  \bibinfo{author}{\bibfnamefont{F.}~\bibnamefont{{Pretorius}}},
  \bibinfo{journal}{ArXiv e-prints}  (\bibinfo{year}{2012}),
  \eprint{1211.4590}.

\bibitem[{\citenamefont{{Siemens} et~al.}(2013)\citenamefont{{Siemens},
  {Ellis}, {Jenet}, and {Romano}}}]{2013arXiv1305.3196S}
\bibinfo{author}{\bibfnamefont{X.}~\bibnamefont{{Siemens}}},
  \bibinfo{author}{\bibfnamefont{J.}~\bibnamefont{{Ellis}}},
  \bibinfo{author}{\bibfnamefont{F.}~\bibnamefont{{Jenet}}}, \bibnamefont{and}
  \bibinfo{author}{\bibfnamefont{J.~D.} \bibnamefont{{Romano}}},
  \bibinfo{journal}{ArXiv e-prints}  (\bibinfo{year}{2013}),
  \eprint{1305.3196}.

\bibitem[{\citenamefont{{Polchinski}}(2007)}]{2007arXiv0707.0888P}
\bibinfo{author}{\bibfnamefont{J.}~\bibnamefont{{Polchinski}}},
  \bibinfo{journal}{ArXiv e-prints}  (\bibinfo{year}{2007}),
  \eprint{0707.0888}.

\bibitem[{\citenamefont{{Allen}}(1997)}]{1997stgr.proc....3A}
\bibinfo{author}{\bibfnamefont{B.}~\bibnamefont{{Allen}}}, in
  \emph{\bibinfo{booktitle}{Some Topics on General Relativity and Gravitational
  Radiation}}, edited by \bibinfo{editor}{\bibfnamefont{J.~A.}
  \bibnamefont{{Miralles}}}, \bibinfo{editor}{\bibfnamefont{J.~A.}
  \bibnamefont{{Morales}}}, \bibnamefont{and}
  \bibinfo{editor}{\bibfnamefont{D.}~\bibnamefont{{Saez}}}
  (\bibinfo{year}{1997}), p.~\bibinfo{pages}{3}.

\bibitem[{\citenamefont{{Dubath} et~al.}(2008)\citenamefont{{Dubath},
  {Polchinski}, and {Rocha}}}]{2008PhRvD..77l3528D}
\bibinfo{author}{\bibfnamefont{F.}~\bibnamefont{{Dubath}}},
  \bibinfo{author}{\bibfnamefont{J.}~\bibnamefont{{Polchinski}}},
  \bibnamefont{and} \bibinfo{author}{\bibfnamefont{J.~V.}
  \bibnamefont{{Rocha}}}, \bibinfo{journal}{\prd}
  \textbf{\bibinfo{volume}{77}}, \bibinfo{eid}{123528} (\bibinfo{year}{2008}),
  \eprint{0711.0994}.

\bibitem[{\citenamefont{{Damour} and {Vilenkin}}(2001)}]{2001PhRvD..64f4008D}
\bibinfo{author}{\bibfnamefont{T.}~\bibnamefont{{Damour}}} \bibnamefont{and}
  \bibinfo{author}{\bibfnamefont{A.}~\bibnamefont{{Vilenkin}}},
  \bibinfo{journal}{\prd} \textbf{\bibinfo{volume}{64}}, \bibinfo{eid}{064008}
  (\bibinfo{year}{2001}), \eprint{arXiv:gr-qc/0104026}.

\bibitem[{\citenamefont{{Damour} and {Vilenkin}}(2005)}]{2005PhRvD..71f3510D}
\bibinfo{author}{\bibfnamefont{T.}~\bibnamefont{{Damour}}} \bibnamefont{and}
  \bibinfo{author}{\bibfnamefont{A.}~\bibnamefont{{Vilenkin}}},
  \bibinfo{journal}{\prd} \textbf{\bibinfo{volume}{71}}, \bibinfo{eid}{063510}
  (\bibinfo{year}{2005}), \eprint{arXiv:hep-th/0410222}.

\bibitem[{\citenamefont{{Abbott} et~al.}(2009)\citenamefont{{Abbott}, {Abbott},
  {Acernese}, {Adhikari}, {Ajith}, {Allen}, {Allen}, {Alshourbagy}, {Amin},
  {Anderson} et~al.}}]{2009Natur.460..990A}
\bibinfo{author}{\bibfnamefont{B.~P.} \bibnamefont{{Abbott}}},
  \bibinfo{author}{\bibfnamefont{R.}~\bibnamefont{{Abbott}}},
  \bibinfo{author}{\bibfnamefont{F.}~\bibnamefont{{Acernese}}},
  \bibinfo{author}{\bibfnamefont{R.}~\bibnamefont{{Adhikari}}},
  \bibinfo{author}{\bibfnamefont{P.}~\bibnamefont{{Ajith}}},
  \bibinfo{author}{\bibfnamefont{B.}~\bibnamefont{{Allen}}},
  \bibinfo{author}{\bibfnamefont{G.}~\bibnamefont{{Allen}}},
  \bibinfo{author}{\bibfnamefont{M.}~\bibnamefont{{Alshourbagy}}},
  \bibinfo{author}{\bibfnamefont{R.~S.} \bibnamefont{{Amin}}},
  \bibinfo{author}{\bibfnamefont{S.~B.} \bibnamefont{{Anderson}}},
  \bibnamefont{et~al.}, \bibinfo{journal}{\nat} \textbf{\bibinfo{volume}{460}},
  \bibinfo{pages}{990} (\bibinfo{year}{2009}), \eprint{0910.5772}.

\bibitem[{\citenamefont{{Boyle} and {Buonanno}}(2008)}]{2008PhRvD..78d3531B}
\bibinfo{author}{\bibfnamefont{L.~A.} \bibnamefont{{Boyle}}} \bibnamefont{and}
  \bibinfo{author}{\bibfnamefont{A.}~\bibnamefont{{Buonanno}}},
  \bibinfo{journal}{\prd} \textbf{\bibinfo{volume}{78}}, \bibinfo{eid}{043531}
  (\bibinfo{year}{2008}), \eprint{0708.2279}.

\bibitem[{\citenamefont{{Smith} et~al.}(2006)\citenamefont{{Smith},
  {Pierpaoli}, and {Kamionkowski}}}]{2006PhRvL..97b1301S}
\bibinfo{author}{\bibfnamefont{T.~L.} \bibnamefont{{Smith}}},
  \bibinfo{author}{\bibfnamefont{E.}~\bibnamefont{{Pierpaoli}}},
  \bibnamefont{and}
  \bibinfo{author}{\bibfnamefont{M.}~\bibnamefont{{Kamionkowski}}},
  \bibinfo{journal}{Phys. Rev. Lett.} \textbf{\bibinfo{volume}{97}},
  \bibinfo{eid}{021301} (\bibinfo{year}{2006}),
  \eprint{arXiv:astro-ph/0603144}.

\bibitem[{\citenamefont{{Archidiacono}
  et~al.}(2013)\citenamefont{{Archidiacono}, {Hannestad}, {Mirizzi}, {Raffelt},
  and {Wong}}}]{2013arXiv1307.0615A}
\bibinfo{author}{\bibfnamefont{M.}~\bibnamefont{{Archidiacono}}},
  \bibinfo{author}{\bibfnamefont{S.}~\bibnamefont{{Hannestad}}},
  \bibinfo{author}{\bibfnamefont{A.}~\bibnamefont{{Mirizzi}}},
  \bibinfo{author}{\bibfnamefont{G.}~\bibnamefont{{Raffelt}}},
  \bibnamefont{and} \bibinfo{author}{\bibfnamefont{Y.~Y.~Y.}
  \bibnamefont{{Wong}}}, \bibinfo{journal}{ArXiv e-prints}
  (\bibinfo{year}{2013}), \eprint{1307.0615}.

\bibitem[{\citenamefont{{Ellis} et~al.}(2011)\citenamefont{{Ellis},
  {McLaughlin}, and {Verbiest}}}]{2011MNRAS.417.2318E}
\bibinfo{author}{\bibfnamefont{J.~A.} \bibnamefont{{Ellis}}},
  \bibinfo{author}{\bibfnamefont{M.~A.} \bibnamefont{{McLaughlin}}},
  \bibnamefont{and} \bibinfo{author}{\bibfnamefont{J.~P.~W.}
  \bibnamefont{{Verbiest}}}, \bibinfo{journal}{\mnras}
  \textbf{\bibinfo{volume}{417}}, \bibinfo{pages}{2318} (\bibinfo{year}{2011}),
  \eprint{1107.4014}.

\bibitem[{\citenamefont{{Hobbs} et~al.}(2006)\citenamefont{{Hobbs}, {Edwards},
  and {Manchester}}}]{2006MNRAS.369..655H}
\bibinfo{author}{\bibfnamefont{G.~B.} \bibnamefont{{Hobbs}}},
  \bibinfo{author}{\bibfnamefont{R.~T.} \bibnamefont{{Edwards}}},
  \bibnamefont{and} \bibinfo{author}{\bibfnamefont{R.~N.}
  \bibnamefont{{Manchester}}}, \bibinfo{journal}{\mnras}
  \textbf{\bibinfo{volume}{369}}, \bibinfo{pages}{655} (\bibinfo{year}{2006}),
  \eprint{arXiv:astro-ph/0603381}.

\bibitem[{\citenamefont{{Estabrook} and
  {Wahlquist}}(1975)}]{1975GReGr...6..439E}
\bibinfo{author}{\bibfnamefont{F.~B.} \bibnamefont{{Estabrook}}}
  \bibnamefont{and} \bibinfo{author}{\bibfnamefont{H.~D.}
  \bibnamefont{{Wahlquist}}}, \bibinfo{journal}{General Relativity and
  Gravitation} \textbf{\bibinfo{volume}{6}}, \bibinfo{pages}{439}
  (\bibinfo{year}{1975}),
  \adsurl{http://adsabs.harvard.edu/abs/1975GReGr...6..439E}.

\bibitem[{\citenamefont{{Favata}}(2010)}]{2010CQGra..27h4036F}
\bibinfo{author}{\bibfnamefont{M.}~\bibnamefont{{Favata}}},
  \bibinfo{journal}{Classical and Quantum Gravity}
  \textbf{\bibinfo{volume}{27}}, \bibinfo{eid}{084036} (\bibinfo{year}{2010}),
  \eprint{1003.3486}.

\bibitem[{\citenamefont{{Seto}}(2009)}]{2009MNRAS.400L..38S}
\bibinfo{author}{\bibfnamefont{N.}~\bibnamefont{{Seto}}},
  \bibinfo{journal}{\mnras} \textbf{\bibinfo{volume}{400}},
  \bibinfo{pages}{L38} (\bibinfo{year}{2009}), \eprint{0909.1379}.

\bibitem[{\citenamefont{{Pshirkov} et~al.}(2010)\citenamefont{{Pshirkov},
  {Baskaran}, and {Postnov}}}]{2010MNRAS.402..417P}
\bibinfo{author}{\bibfnamefont{M.~S.} \bibnamefont{{Pshirkov}}},
  \bibinfo{author}{\bibfnamefont{D.}~\bibnamefont{{Baskaran}}},
  \bibnamefont{and} \bibinfo{author}{\bibfnamefont{K.~A.}
  \bibnamefont{{Postnov}}}, \bibinfo{journal}{\mnras}
  \textbf{\bibinfo{volume}{402}}, \bibinfo{pages}{417} (\bibinfo{year}{2010}),
  \eprint{0909.0742}.

\bibitem[{\citenamefont{{Jenet}
  et~al.}(2005{\natexlab{b}})\citenamefont{{Jenet}, {Creighton}, and
  {Lommen}}}]{2005ApJ...627L.125J}
\bibinfo{author}{\bibfnamefont{F.~A.} \bibnamefont{{Jenet}}},
  \bibinfo{author}{\bibfnamefont{T.}~\bibnamefont{{Creighton}}},
  \bibnamefont{and} \bibinfo{author}{\bibfnamefont{A.}~\bibnamefont{{Lommen}}},
  \bibinfo{journal}{\apjl} \textbf{\bibinfo{volume}{627}},
  \bibinfo{pages}{L125} (\bibinfo{year}{2005}{\natexlab{b}}),
  \eprint{arXiv:astro-ph/0505585}.

\end{thebibliography}

%\end{thebibliography}

\end{document}